\documentclass[aps, pra, twocolumn, amsmath, amssymb, longbibliography, superscriptaddress, floatfix, final]{revtex4-2}

\usepackage{graphicx, dcolumn, hhline, bm, amsmath, amsfonts, amssymb, color, tabularx, comment}
\usepackage[caption=false]{subfig}
\usepackage{floatrow}
\usepackage{physics}
\usepackage{appendix}
\usepackage{adjustbox}
\usepackage{xcolor}
\usepackage[%
  pdftex,
  breaklinks=true,
  colorlinks=true,
  linkcolor=blue,
  urlcolor=blue,
  citecolor=blue
]{hyperref}

\newcommand{\NM}[1]{\color{black} #1 \color{black}}

\begin{document}
\title{Resonant squeezed light from photonic Cooper pairs}

\author{Sanker Timsina}
\affiliation{Department of Physics and Astronomy, University of Victoria, Victoria, British Columbia V8W 2Y2, Canada}
\affiliation{Centre for Advanced Materials and Related Technology, University of Victoria, Victoria, British Columbia V8W 2Y2, Canada}

\author{Taha \surname{Hammadia}}
\affiliation{Department of Physics and Astronomy, University of Victoria, Victoria, British Columbia V8W 2Y2, Canada}
\affiliation{Centre for Advanced Materials and Related Technology, University of Victoria, Victoria, British Columbia V8W 2Y2, Canada}
\affiliation{\'{E}cole Polytechnique, Institut Polytechnique de Paris, 91120 Palaiseau, France}

\author{Sahar \surname{Gholami Milani}}
\affiliation{Department of Physics and Astronomy, University of Victoria, Victoria, British Columbia V8W 2Y2, Canada}
\affiliation{Centre for Advanced Materials and Related Technology, University of Victoria, Victoria, British Columbia V8W 2Y2, Canada}

\author{Filomeno S. \surname{de Aguiar J\'{u}nior}}
\affiliation{Department of Chemistry, University of Victoria, Victoria, British Columbia V8W 2Y2, Canada}
\affiliation{Centre for Advanced Materials and Related Technology, University of Victoria, Victoria, British Columbia V8W 2Y2, Canada}

\author{Alexandre \surname{Brolo}}
\affiliation{Department of Chemistry, University of Victoria, Victoria, British Columbia V8W 2Y2, Canada}
\affiliation{Centre for Advanced Materials and Related Technology, University of Victoria, Victoria, British Columbia V8W 2Y2, Canada}

\author{Rog\'{e}rio \surname{de Sousa}}
\email[Corresponding author: ]{rdesousa@uvic.ca}
\affiliation{Department of Physics and Astronomy, University of Victoria, Victoria, British Columbia V8W 2Y2, Canada}
\affiliation{Centre for Advanced Materials and Related Technology, University of Victoria, Victoria, British Columbia V8W 2Y2, Canada}

\date{\today}

\begin{abstract}
Raman scattering of photons into phonons gives rise to entangled photon pairs when the phonon emitted in a Stokes process is coherently absorbed in antiStokes scattering, forming the photonic analog of Cooper pairs. We present a nonperturbative theory for the time evolution of photonic Cooper pairs that treats interacting photons and phonons as a hybrid excitation, the Ramaniton. 
As the Ramaniton propagates in a wave guide it displays quantum oscillations between photon and phonon occupation, leading to resonant squeezed Stokes-antiStokes light when the phonon occupation becomes equal to zero \NM{without recurring back to the photon vacuum}. 
This phenomenon is predicted to generate up to 28 dB of squeezed light even in standard silicon on insulator waveguides.
\end{abstract}

\maketitle

\begin{figure}
\centering
\includegraphics[width=\linewidth]{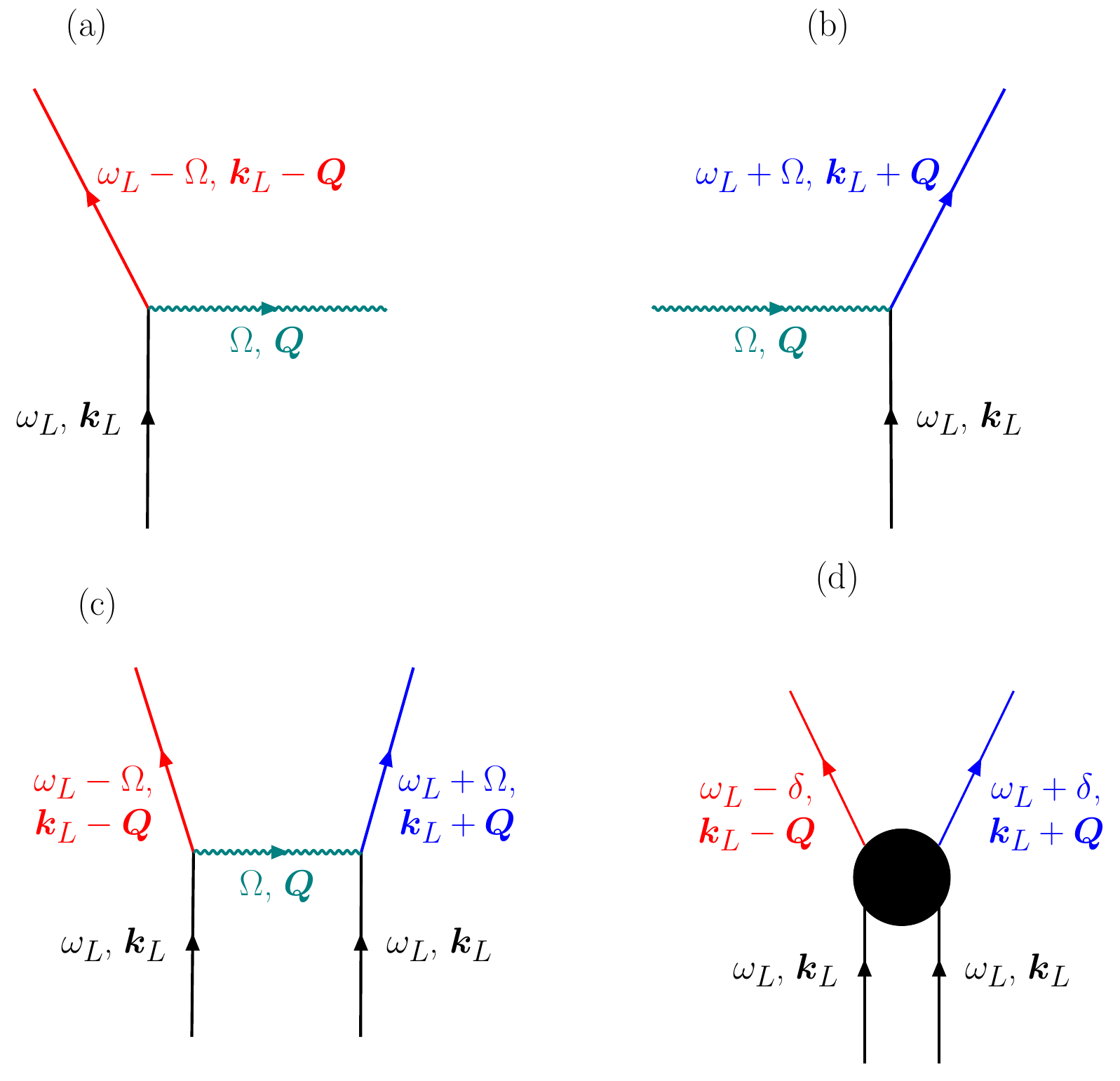}
\caption{Two and four-photon Raman scattering processes. (a) Stokes: An incident laser photon with frequency $\omega_L$ and momentum $\bm{k}_L$ scatters into a red shifted Stokes photon with frequency $\omega_L-\Omega$ and momentum $\bm{k}_L-\bm{Q}$, and emits a phonon with frequency $\Omega$ and momentum $\bm{Q}$. (b) anti-Stokes process: The laser photon absorbs a phonon and emits a blue shifted photon with frequency $\omega_L+\Omega$ and momentum $\bm{k}_L+\bm{Q}$. (c) \NM{Formation of Stokes-antiStokes pair mediated by a ``real'' phonon:} The phonon emitted by a Stokes event is coherently absorbed by another incident laser photon, leading to an entangled Stokes-antiStokes (SaS) pair. (d) \NM{Formation of Stokes-antiStokes pair mediated by ``virtual'' excitations to phonon or electron orbital states: In this case the frequency of the Stokes/antiStokes photons can be $\omega_L\pm \delta$ with $\delta\neq \Omega$. When the phonon resonance condition $\delta=\Omega$ is satisfied, the ``real'' diagram (c) is expected to dominate the formation of Stokes-antiStokes pairs. This phonon resonance happens when the normalized Raman shift $q=\delta/\Omega=c'Q/\Omega=1$, where $c'$ is the speed of light in the material}.}
\label{fig:SaSDiagram}
\end{figure}

\section{Introduction}

A key challenge for the development of quantum photonic technology is the efficient generation of photon entanglement in solid-state chips \cite{Wang2020}. Low noise generation of entangled photon pairs is a necessary condition for several applications, ranging from quantum sensing \cite{Schnabel2017} and quantum computing \cite{Fukui2018} with squeezed states to distributed quantum computing in a future quantum internet \cite{Wehner2018}. 
Realizing these applications requires maximization of photon entanglement \cite{Duan2000}, together with minimization of noise from photon loss and thermal emission in the solid-state environment. To achieve this, it is desirable to develop microscopic models that account for light-matter interaction beyond the usual free-space quantum optic phenomenology. 

A recent development was the realization that vibrational modes of molecules and crystals (phonons) can act as mediators for photon-photon interaction in a wide variety of substances \cite{Kasperczyk2016, Velez2020}.  The origin of this phenomenon is a correlated Raman process, whereby a phonon emitted by a Stokes photon scattering event is coherently absorbed by another incident photon, generating a Stokes-antiStokes (SaS) photon pair, see Fig.~\ref{fig:SaSDiagram}. This phenomenon is the photon analogue of the attractive interaction that forms Cooper pairs in superconductors, leading to the idea that the SaS state is a ``photonic Cooper pair'' \cite{Saraiva2017}.

So far experimental and theoretical studies of photonic Cooper pairs have focused solely on the regime of short time evolution or scattering \cite{Guimaraes2020}, and the question of what photonic state emerges for longer time evolution in a waveguide is open. As suggested in Figs.~\ref{fig:SaSDiagram}(c, d), photonic Cooper pair formation provides a microscopic mechanism for spontaneous four-wave mixing; however, it is not known \NM{how to distinguish or exploit the phonon resonance depicted in Fig.~\ref{fig:SaSDiagram}(c)}
from the usual photon-photon interaction mediated by higher-energy virtual transitions to electronic orbital states of the material \cite{Boyd2020}. 
The modelling of photonic devices traditionally assumes phenomenological quartic-in-electric-field interactions to describe spontaneous four-wave mixing \cite{Banic2022}; it is not known whether the dominant mechanism is of phononic or electronic origin, \NM{and whether it is ``virtual'' or ``real''}.

Here we propose a nonperturbative theory for phonon-mediated generation of entangled photon pairs in Raman scattering. The theory introduces the concept of the Ramaniton, a hybrid photon-phonon excitation that is different from the polariton \cite{Kittel1987} in at least two ways: First, Ramanitons have zero electric dipole moment and do not cause photon loss \cite{Diniz2020}; and second, the excited states of Ramanitons are made of superpositions of photons/phonons as well as their corresponding antiparticles (holes), a direct consequence of the presence of Cooper-pair correlations. 
Interpreting the photonic excitations inside the material as Ramanitons allows the prediction of the quantum state of light that arises in a gas of photonic Cooper pairs. Our theory demonstrates \emph{resonant squeezing} for certain photon propagation lengths, leading to applications in the optimization of quantum photonic devices.

\section{Model for light-matter interaction mediated by a Raman phonon}

We consider a model for Stokes (S) and antiStokes (aS) photons interacting with an optical phonon, the THz lattice vibration of a crystal. Their noninteracting Lagrangian density is given by \cite{Kittel1987, Jackson1998}
\begin{eqnarray}
    {\cal L}_0 &=& \frac{\epsilon}{2} \left( \left|\bm{E}_{S}\right|^{2} + \left|\bm{E}_{aS}\right|^{2}\right) - \frac{1}{2 \mu} \left(\left|\bm{B}_{S}\right|^{2} + \left| \bm{B}_{aS}\right|^{2} \right)\nonumber\\
    &&+ \frac{\rho}{2} \left|\dot{u}\right|^{2} - \frac{\rho}{2} \Omega^2 \left| u\right| ^{2},
    \label{L0}
\end{eqnarray}
where vectors $\bm{E}_j=-\dot{\bm{A}}_j$ and $\bm{B}_j=\nabla\times \bm{A}_j$, $j=S, aS$, are the electric and magnetic fields for the Stokes (S) and antiStokes (aS) photons, which are written in terms of vector potentials $\bm{A}_j$. The scalar $u$ models a phonon displacement with mass density $\rho$ and frequency $\Omega$ independent of wave vector $\bm{Q}$ (a dispersionless mode, describing the flat band of optical phonons near $Q=0$).

We assume $u$ in Eq.~(\ref{L0}) is a \emph{pure} Raman mode, i.e. one that is invariant under the inversion symmetry operation that takes $\bm{E}_j$ into $-\bm{E}_j$. This necessarily implies that the lowest order interaction between $u$ and photons is \NM{linear in phonon displacement and quadratic in electric fields,
\begin{equation}
    {\cal L}_{{\rm int}} = \frac{\xi}{2}u E_{L,x} \left(E_{S,y} +E_{aS,y}\right) + {\rm c.c.},
\label{Lint}
\end{equation}
where $\xi$ is a coupling constant and $E_{L,x}$ is the amplitude of a pump laser polarized along the $x$ direction. The particular choice of interaction (\ref{Lint}) is allowed by the point group symmetry of the diamond lattice, therefore it is exact for crystalline silicon and diamond}. 
Similar interactions are also 
present in 
associated amorphous dielectrics SiO$_x$ and SiN$_{x}$ \cite{Wang2020}, where phonons are either pure infrared ($u$ changes sign under inversion) or pure Raman ($u$ does not change sign under inversion). While materials without an inversion center have phonons with mixed symmetry, interactions like (\ref{Lint}) are still present in them provided we interpret $u$ as the component of a phonon displacement that does not change sign under inversion. That is the part of the phonon that does not have electric dipole moment and as a consequence does not cause photon loss \cite{Diniz2020}.
  
Assuming the pump power is much larger than the S, aS output allows us to take the ``parametric approximation'' and replace $E_{L,x}$ by a complex classical field, $E_{L,x}=E_L e^{i(k_L z-\omega_L t)}$ where $E_L$ is the pump amplitude, and $\omega_L=c'k_L$ is the frequency of the laser, with $c' = 1/\sqrt{\epsilon \mu}=c/n_0$ the speed of light in the material. 

The quantum Hamiltonian is obtained from the total Lagrangian density ${\cal L}={\cal L}_0+{\cal L}_{{\rm int}}$ by assuming the dynamical variables $\bm{A}_S$ and $\bm{A}_{aS}$ are quantum operators with usual commutation relations with their canonical momenta $\partial {\cal L}/\partial \dot{\bm{A}}_j$. After going to $k$ space we get the following Hamiltonian,
\begin{widetext}
\begin{eqnarray}
        {\cal H} &=& \hbar \sum_{\bm{Q}} \left[\omega_{\bm{k}_L + \bm{Q}} \left(b^{\dag}_{\bm{Q}} b_{\bm{Q}} + \frac{1}{2}\right)+\omega_{\bm{k}_L - \bm{Q}} \left(b^{\dag}_{-\bm{Q}} b_{-\bm{Q}} + \frac{1}{2}\right)+ \Omega \left(c^{\dag}_{\bm{Q}} c_{\bm{Q}} + \frac{1}{2}\right)\right.\nonumber\\
        &&\left.+ i \frac{\eta}{2} \cos(\omega_L t) \sqrt{\Omega \omega_{\bm{k}_L + \bm{Q}}} \left(c^\dagger_{\bm{Q}} + c_{\bm{Q}}\right) b^\dagger_{\bm{Q}}+ i \frac{\eta}{2} \cos(\omega_L t) \sqrt{\Omega \omega_{\bm{k}_L - \bm{Q}}} \left(c^\dagger_{\bm{Q}} + c_{\bm{Q}}\right) b^\dagger_{-\bm{Q}} + {\rm H.c.}\right],
\label{Hcomplete}
\end{eqnarray}
\end{widetext}
where $b^{\dag}_{\pm\bm{Q}}, b_{\pm\bm{Q}}$ are creation and annihilation operators for photons \NM{polarized along $y$} with momentum $\bm{k}_L\pm\bm{Q}$ and frequency $\omega_{\bm{k}_L\pm \bm{Q}}=c'|\bm{k}_L\pm\bm{Q}|$; $c^{\dag}_{\bm{Q}}, c_{\bm{Q}}$ are creation and annihilation operators for the phonon mode. The dimensionless parameter $\eta = \xi E_L/(\Omega\sqrt{\rho \epsilon})$ is the figure of merit for the strength of the photon-phonon coupling; it plays a key role in our theory below. 

We move to a rotating frame by applying the unitary transformation $e^{i \omega_L t \left(b^{\dag}_{\bm{Q}} b_{\bm{Q}} + b^{\dag}_{-\bm{Q}} b_{-\bm{Q}}\right)}$, and take the rotating wave approximation by dropping the terms in Eq.~(\ref{Hcomplete}) that 
are never able to
conserve energy, $c_{\bm{Q}} b^{\dag}_{-\bm{Q}}$ and $c^{\dag}_{\bm{Q}} b^{\dag}_{\bm{Q}}$. These terms lead to time-oscillatory contributions with high frequency and small amplitude for all observables of interest. Also note that phase matching is trivial in the absence of dispersion in $c'$. Without loss of generality we focus on the case of forward scattering where $\bm{Q}\parallel \bm{k}_L$; this implies the mode $b_{-\bm{Q}}\equiv b_{S}$ is red shifted in frequency, and  $b_{\bm{Q}}\equiv b_{aS}$ is blue shifted. Altogether the approximations lead to a time-independent Hamiltonian that we call the Ramaniton model:
\begin{eqnarray} 
    {\cal H}_{{\rm R}} &=& \hbar\Omega\left[-q b^{\dag}_{S} b_{S} + q b^{\dag}_{aS} b_{aS} + c^\dagger c + \frac{1}{2} \right.\nonumber\\ 
    &&+\left. i \eta_- \left(c^\dagger b^\dagger_S - c b_S\right) +  i \eta_+ \left(c b^\dagger_{aS} - c^\dagger b_{aS}\right)\right],
\label{RamanitonModel}
\end{eqnarray}
where we wrote the dimensionless Raman shift as $q = \frac{c' Q}{\Omega}$, \NM{with $0 \le q\le \omega_L/\Omega$, and the dimensionless coupling constants as 
\begin{equation}
    \eta_{\pm} = \frac{\eta}{4} \sqrt{\left(\frac{\omega_L}{\Omega}\right) \pm q},
\end{equation}
with $\eta = \xi E_L/(\Omega\sqrt{\rho \epsilon})$ the figure of merit for light-matter interaction}. 
Hamiltonian (\ref{RamanitonModel}) provides the microscopic realization of the phenomenological model proposed in \cite{Saraiva2017}. This shows that the degree of light-matter interaction set by $\eta_{\pm}$ can be tuned by the pump laser frequency $\omega_L$ and amplitude $E_L$ via $\eta$.

As a first attempt to tackle the phonon-mediated interaction we follow Saraiva {\it et al.}  \cite{Saraiva2017} and apply the Schrieffer-Wolff canonical transformation to Eq.~(\ref{RamanitonModel}). This converts the terms linear in $\eta_{\pm}$ to a Cooper-pair like interaction,
\begin{equation}
{\cal H}_{{\rm S-W}}=  \frac{\hbar\Omega\eta_{+}\eta_{-}}{1-q}\left(b^{\dag}_{S}b^{\dag}_{aS}+b_{aS}b_{S}\right),
\label{S-Wint}
\end{equation}
plus a series of additional higher-order terms such as $\eta_{-}^{3}/(1-q)^2 b^{\dag}_{S}c^{\dag}$. 


If we assume the additional higher-order terms can be neglected, the time evolution of photonic Cooper pairs from the vacuum at $t=0$ is predicted to be
\begin{equation}
e^{-i\frac{t}{\hbar}{\cal H}_R}\ket{0,0,0}\approx 
e^{-i\frac{\Omega t}{2}}\sum_{N=0}^{\infty} \frac{(-i)^{N}\tanh^{N}{(r)}}{\cosh{(r)}}\ket{N,0,N}, 
\label{psiSW}
\end{equation}
where $r=\eta_{+}\eta_{-}(\Omega t)/(1-q)$ and $\ket{N_S,N_c, N_{aS}}$ denotes number states of Stokes, phonons, and antiStokes, respectively. 

The state (\ref{psiSW}) is ``photon paired''; it is known to generate two-mode squeezed states of light \NM{that are entangled \cite{Duan2000}
and} play a key role in quantum sensing and computing \cite{Gerry2004, Walls2008}. However, the Schrieffer-Wolff method leads to a perturbative series in powers of $\eta/(1-q)$. The series blows up when $q\rightarrow 1$, the phonon resonance condition for ``real'' emission/absorption of a phonon. In this regime Eqs.~(\ref{S-Wint})~and~(\ref{psiSW}) are not expected to be good approximations. 
Below we propose a nonperturbative solution to this problem. 

\begin{figure}
\centering
\includegraphics[width=\linewidth]{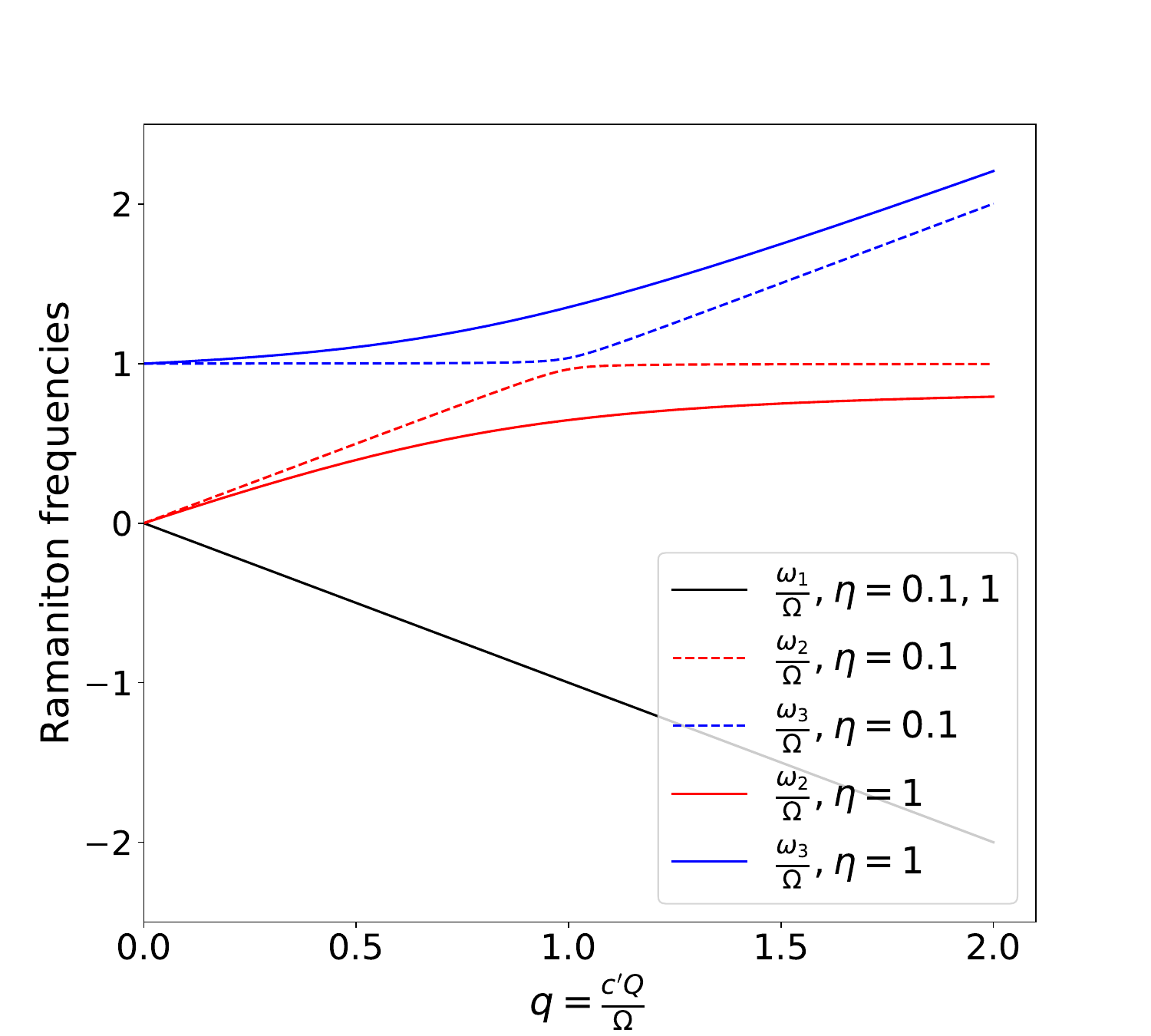}
\caption{Ramaniton dispersions for $\omega_L/\Omega=12.4$, $\eta=0.1$ (dashed lines), and $\eta=1$ (solid lines). The $\omega_1$ Ramaniton is decoupled from the phonon and is independent of $\eta$. When $\eta \ll 1$, photon-phonon hybridization is significant only in the neighbourhood of $q=1$ (phonon resonance), where there is avoided crossing between $\omega_2$ and $\omega_3$. In contrast, for larger $\eta$ the modes $\omega_2, \omega_3$ are hybridized in a wider range of $q$. 
\label{fig:RamanitonDispersion}}
\end{figure}

\section{Nonperturbative theory\label{sec:nonpert}}

Here we propose instead to diagonalize the Ramaniton model exactly using the Nambu method, required to find quantum excitations of superconducting \cite{Nambu1960} and antiferromagnetic materials \cite{Beairsto2021}. In the Nambu representation the Hamiltonian (\ref{RamanitonModel}) is written as 
\begin{equation}
    {\cal H}_R = \frac{1}{2} \bm{v}^{\dag} \cdot\mathbb{L}\cdot \bm{v},
\label{H_RNambu}
\end{equation}
where $\bm{v}=\left(b_S, c, b_{aS}, b^{\dag}_S, c^{\dag}, b^{\dag}_{aS} \right)^T$ is a particle-antiparticle column vector, and $\mathbb{L}$ is a $6\times 6$ Hermitian matrix.
The goal of the Nambu method is to find the set of operators $\bm{\alpha} = \mathbb{U}^{-1} \cdot\bm{v}$ that leads to
\begin{equation} 
\label{diagonal-ham}
    {\cal H}_R = \sum_{j = 1}^{3} \hbar \omega_j \left(\alpha^{\dag}_j \alpha_j + \frac{1}{2}\right). 
\end{equation}
To see how this is done, first note that in order to preserve the Bosonic commutation relations for the new operators $\bm{\alpha}$ we need to use a nonunitary canonical transformation $\mathbb{U}$ that satisfies $\mathbb{U}^{-1}=\mathbb{Z}\cdot\mathbb{U}^{\dag}\cdot\mathbb{Z}$,
where $\mathbb{Z}={\rm diag}\{1,1,1,-1,-1,-1 \}$ is the plus (minus) identity in the particle (antiparticle or hole) subspace. 
From Eq.~(\ref{diagonal-ham}) we get $\left[\bm{\alpha},{\cal H}_R\right]=\mathbb{Z}\cdot \mathbb{W}\cdot \bm{\alpha}$, where $\mathbb{W}={\rm diag}\{\omega_1,\omega_2,\omega_3,\omega_1,\omega_2,\omega_3\}$. Compare to $\left[\mathbb{U}^{-1}\cdot \bm{v},{\cal H}_R\right]=\mathbb{U}^{-1}\cdot\mathbb{Z}\cdot\mathbb{L}\cdot\bm{v}$ to get $(\mathbb{Z}\cdot\mathbb{L})\cdot\mathbb{U}=\mathbb{U}\cdot(\mathbb{Z}\cdot\mathbb{W})$, showing that the columns of $\mathbb{U}$ are the right eigenvectors of $\mathbb{Z}\cdot\mathbb{L}$. \NM{The first three columns have eigenvalue $+\omega_j$ and the last three have $-\omega_j$}. Thus the Nambu method is to diagonalize the non-Hermitian matrix $\mathbb{Z}\cdot\mathbb{L}$, use its eigenvectors to form the columns of $\mathbb{U}$, and make sure the eigenvector normalization satisfies 
$(\mathbb{Z}\cdot\mathbb{U}^{\dag}\cdot\mathbb{Z})\cdot \mathbb{U}=\mathbb{I}$, where $\mathbb{I}$ is the identity matrix. \NM{The details of this procedure are described in Appendix~\ref{appendixNambu}}.

\begin{figure*}
\begin{center}
\subfloat[]{\includegraphics[width=.5\linewidth]{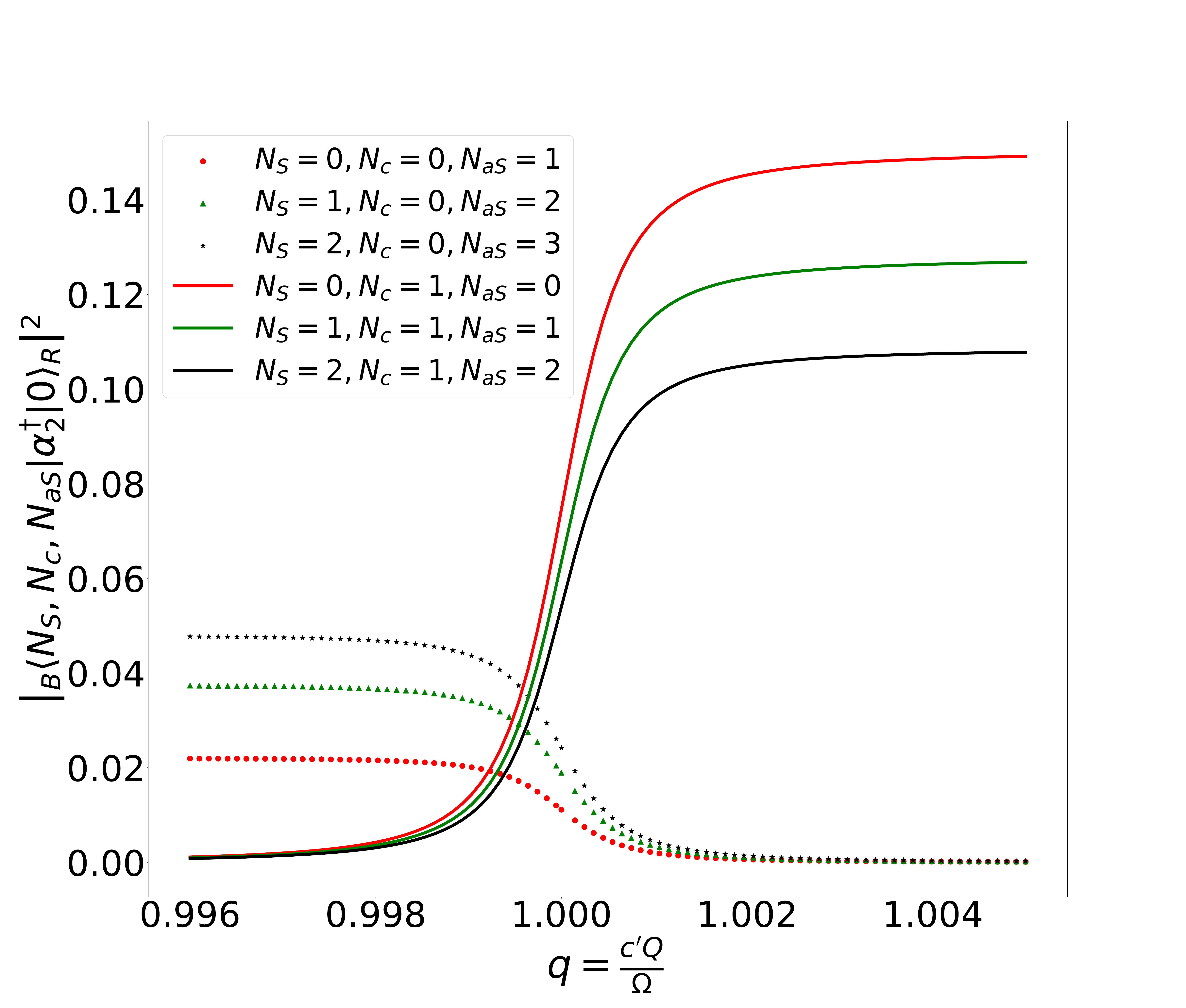}}
\subfloat[]{\includegraphics[width=.5\linewidth]{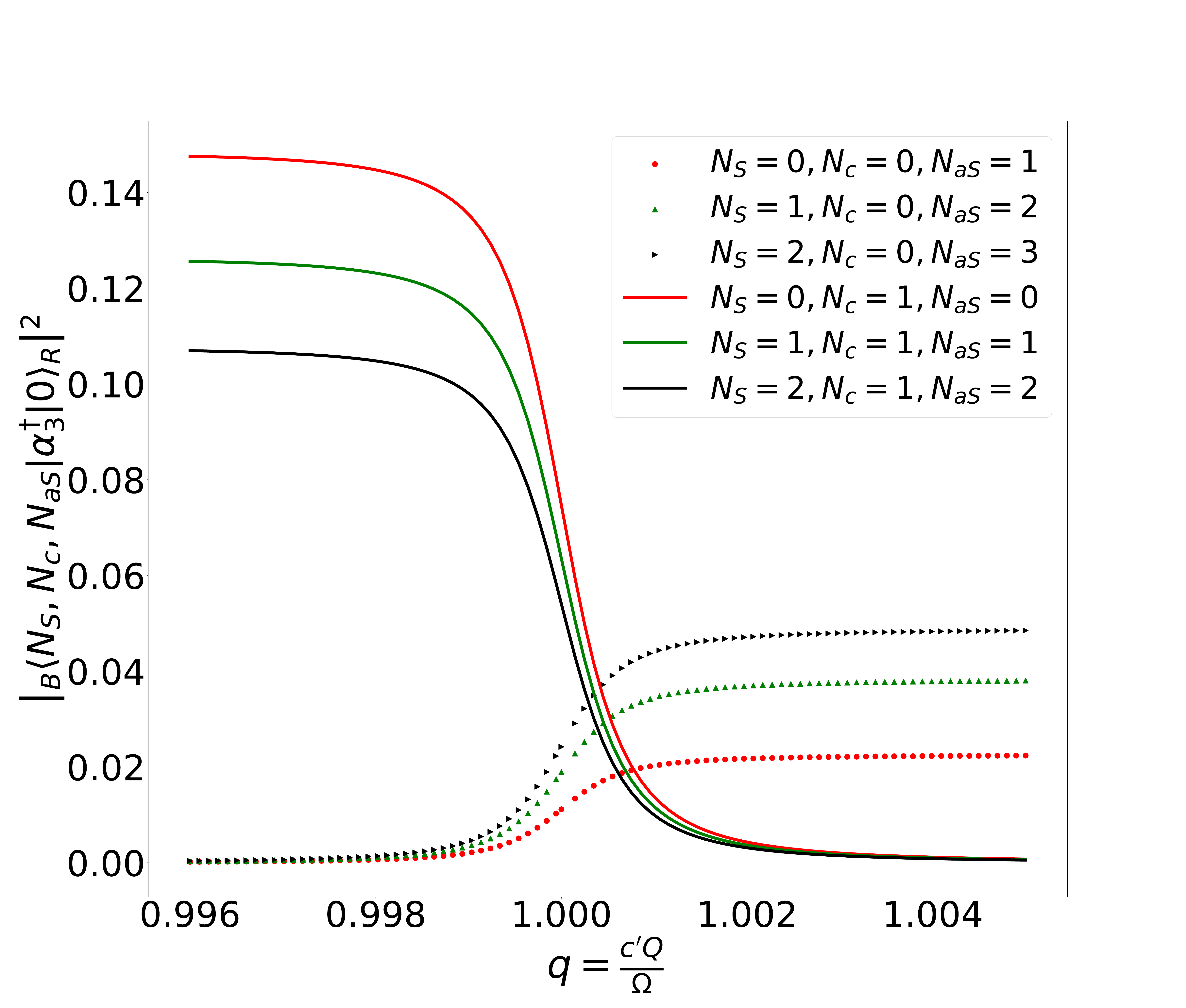}}
\end{center}
\caption{\NM{Bare photon and phonon probabilities for singly-occupied Ramaniton states (a) $\alpha^{\dag}_{2}\ket{0}_R$ and (b) $\alpha^{\dag}_{3}\ket{0}_R$ ($\ket{0}_R$ is the Ramaniton vacuum). The parameters are realistic for silicon chips, $\omega_L/\Omega=12.4$ and $\eta=10^{-3}$. These states contain bare modes $\ket{N_S,N_c,N_{aS}}=\ket{N,0,N}_B$ and $\ket{N,1,N}_B$ for $N=0,1,2,3,\ldots$. As $q$ is increased over the phonon resonance at $q=1$ the character of (a) changes from phonon-poor to phonon-rich, with (b) showing the opposite behaviour. \label{fig:bare_prob23}}}
\end{figure*}

The operators $\alpha^{\dag}_j$ create the hybrid phonon-photon excitations that we call Ramanitons. \NM{Exact diagonalization of $\mathbb{Z}\cdot\mathbb{L}$ leads to 6 eigenvalues, the diagonal elements of $\mathbb{Z}\cdot\mathbb{W}$: $\omega_1, \omega_2, \omega_3, -\omega_1, -\omega_2, -\omega_3$. The choice of which ones are taken as positive or negative is strictly determined by the canonical transformation constraint $(\mathbb{Z}\cdot\mathbb{U}^{\dag}\cdot\mathbb{Z})\cdot \mathbb{U}=\mathbb{I}$ as explained in Appendix~\ref{appendixNambu}}. These are the Ramaniton dispersion relations,
\begin{subequations}
\begin{eqnarray}
    \omega_1 &=& -q\Omega,\label{omega1}\\
    \omega_2 &=& \left(\frac{q+1}{2} - \frac{1}{2} \sqrt{(q-1)^2 + \frac{\eta^2 q}{2}} \right) \Omega,\label{omega2}\\
    \omega_3 &=& \left(\frac{q +1}{2} + \frac{1}{2} \sqrt{(q - 1)^2 + \frac{\eta^2 q}{2}} \right)\Omega.\label{omega3}
\end{eqnarray}
\end{subequations}
Some of the Ramaniton frequencies are negative in the rotating frame; as a check, note that taking the limit $\eta\rightarrow 0$ leads to frequencies $\pm q\Omega=\pm c'Q$ and $\Omega$, consistent with Eq.~(\ref{RamanitonModel}) when $\eta=0$. 

The creation operator for the Ramaniton with frequency $\omega_1$ is given by
\begin{equation}
\alpha^{\dag}_{1}=\frac{\eta_+}{\sqrt{\eta^{2}_{+}-\eta^{2}_{-}}}b^{\dag}_{S} - \frac{\eta_-}{\sqrt{\eta^{2}_{+}-\eta^{2}_{-}}}b_{aS}.
\label{alpha1bare}
\end{equation}
This shows that $\omega_1$ is a pure photon mode that is \emph{decoupled} from the phonon. It is given by the quantum superposition of ``particle-like'' Stokes and ``hole-like'' antiStokes photons. The latter corresponds to the absence of an antiStokes excitation, with frequency equal to minus the aS frequency, $-(+c'Q)$, which turns out to be the same as the Stokes frequency. 
Antiparticle or hole-like states appear because the vacuum of the Ramaniton is ``dressed'' with excitations. 
The other excitations $\omega_2, \omega_3$ are hybrid in that they mix (anti)photons and (anti)phonons \NM{as can be seen in Eqs.~(\ref{alpha2}),~(\ref{alpha3})}. They lead to avoided-crossing of photon and phonon dispersions, with maximum hybridization happening close to the phonon resonance at $q=1$. The laser amplitude $E_L$ is able to modify these modes by increasing $\eta$. The dispersions are shown in Fig.~\ref{fig:RamanitonDispersion}.

\NM{The Ramaniton number states are formed by superpositions of multiple bare photons and phonon number states. The vacuum for the Ramanitons is given by
\begin{equation}
\label{RVS2}
\ket{0}_R= \sqrt{1 - \left(\frac{\eta_-}{\eta_+}\right)^2}\sum_{N=0}^{\infty} \left(\frac{\eta_-}{\eta_+}\right)^N  \ket{N,0,N}_B,
\end{equation}
where $\ket{N_S,N_c,N_{aS}}_B$ are Fock states for the bare modes. Hence, the vacuum of the Ramanitons is a two-mode squeezed state for S and aS photons.
The bare photon probabilities for the states $\alpha_{j}^{\dag}\ket{0}_R$ with $j=2$ and $j=3$ are shown in Figs.~\ref{fig:bare_prob23}(a)~and~\ref{fig:bare_prob23}(b), respectively. As $q$ increases from below to above phonon resonance, the state $\alpha_{2}^{\dag}\ket{0}_R$ changes from multiple photons with nearly zero phonons ($q<1$) to multiple photons with nearly one phonon ($q>1$). The opposite behaviour is observed for  $\alpha_{3}^{\dag}\ket{0}_R$. The change in photon-phonon hybridization across the anticrossing point at $q=1$ is also found in simple phonon-polaritons \cite{Kittel1987}. 

Below the anticrossing at $q\ll 1$}, Eq.~(\ref{omega2}) becomes $\omega_2 \approx (1 - \frac{\eta^2}{8}) c' Q$, showing that the index of refraction depends on the laser intensity $I = c' \epsilon E_L^2/2$ 
according to $n \approx n_0 (1 + \frac{\eta^2}{8}) \equiv n_0 + n_2 I$ in the low $I$ regime. This shows that $\eta$ contributes to the Kerr effect (intensity dependence of the index of refraction), giving rise to a nonlinear index of refraction $n_2$ via the relation $\eta = \sqrt{8 n_2 I/n_0}$. This observation can be used to estimate $\eta$ in photonic materials/devices. 

Raman spectra of bulk silicon shows a single strong sharp resonance \NM{due to three degenerate optical phonons $(u_{yz},u_{xz},u_{xy})$ at $\Omega/2\pi=15.6$~THz  \cite{Hart1970}. 
Only one of them ($u\equiv u_{xy}$) couples to a pump laser polarized along $x$ and propagating along $z$; this generates SaS pairs and Ramaniton states with bare photon polarization along $y$.  
If the pump is near} the $1550$~nm telecom wavelength ($\omega_L/2\pi=193$~THz), photon loss will be so small it can hardly be detected \cite{Schinke2015}. This supports a model that neglects photon loss into channels other than the Raman phonon. For this regime $\omega_L/\Omega=12.4$, with measurements of $n_0=3.42$, and $n_2=4.5\times 10^{-18}$~m$^2/$W \cite{Leuthold2010}. Using  intensities achievable on chip ($I=10^{11}$~W/m$^{2}$ \cite{Vaidya2020}) we estimate $\eta\approx 10^{-3}$ for common silicon waveguides. 

\begin{figure}
\centering
\includegraphics[width=\linewidth]{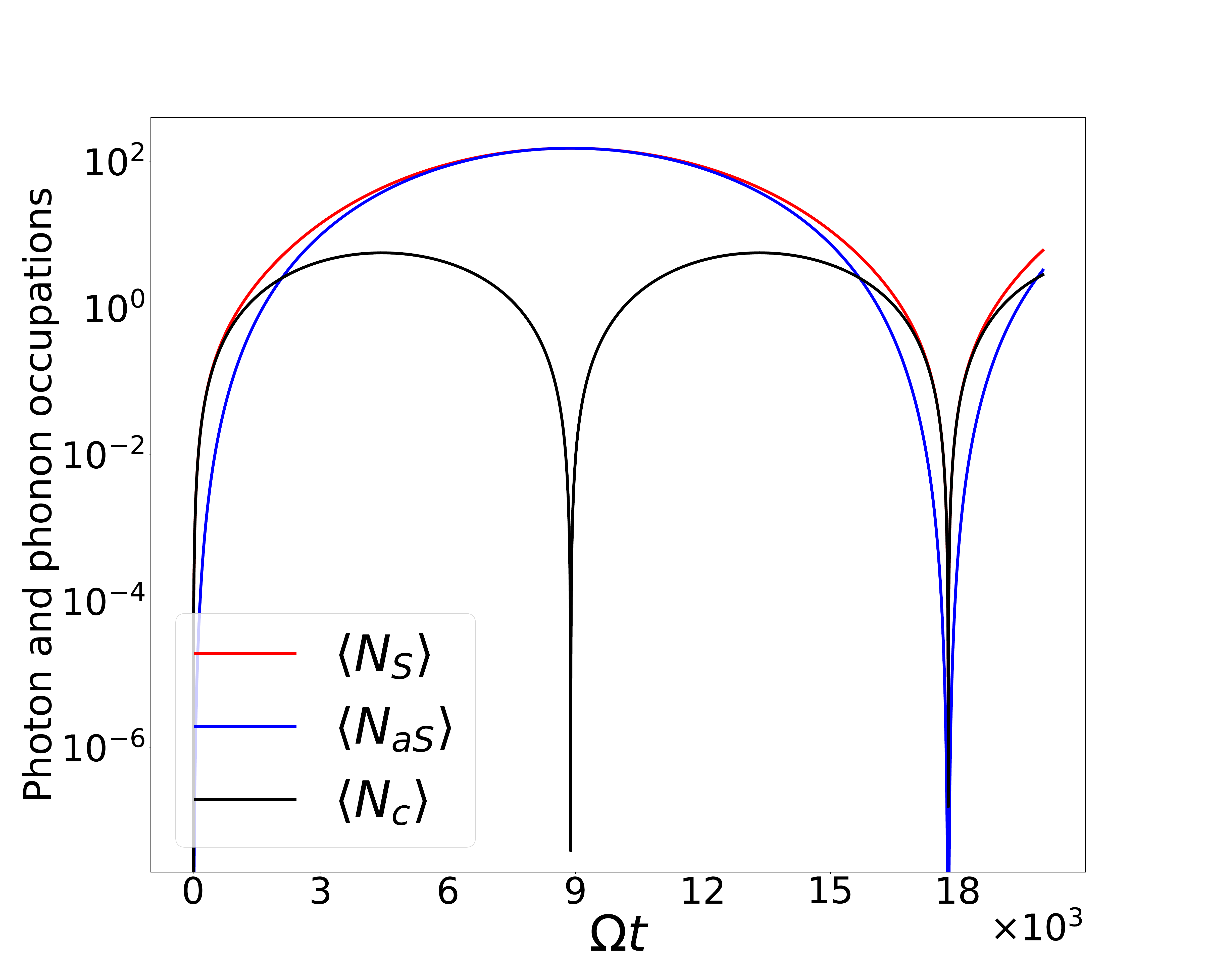}
\caption{Time dependence of occupations for phonon $\langle N_c(t)\rangle$, photon Stokes $\langle N_S(t)\rangle$, and antiStokes $\langle N_{aS}(t)\rangle$ when their initial state is the vacuum, $|\psi(t=0)\rangle= |0\rangle_S |0\rangle_{c} |0\rangle_{aS}$, and the phonon resonance condition is satisfied ($q=1$). The parameters are realistic for silicon chips, $\omega_L/\Omega=12.4$ and $\eta=10^{-3}$. 
The occupations satisfy the symmetry $\langle N_S(t)\rangle-\langle N_{aS}(t)\rangle=\langle N_c(t)\rangle$ so that $\langle N_S(t)\rangle\geq \langle N_{aS}(t)\rangle,\langle N_c(t)\rangle$ for all $t$. 
When $\langle N_c(t)\rangle=0$ (at e.g. $\Omega t=8.89\times 10^3$), the SaS state is ``paired up'' giving rise to maximum squeezing.
\label{fig:occupation}}
\end{figure}

\begin{figure}
\centering
\includegraphics[width=\linewidth]{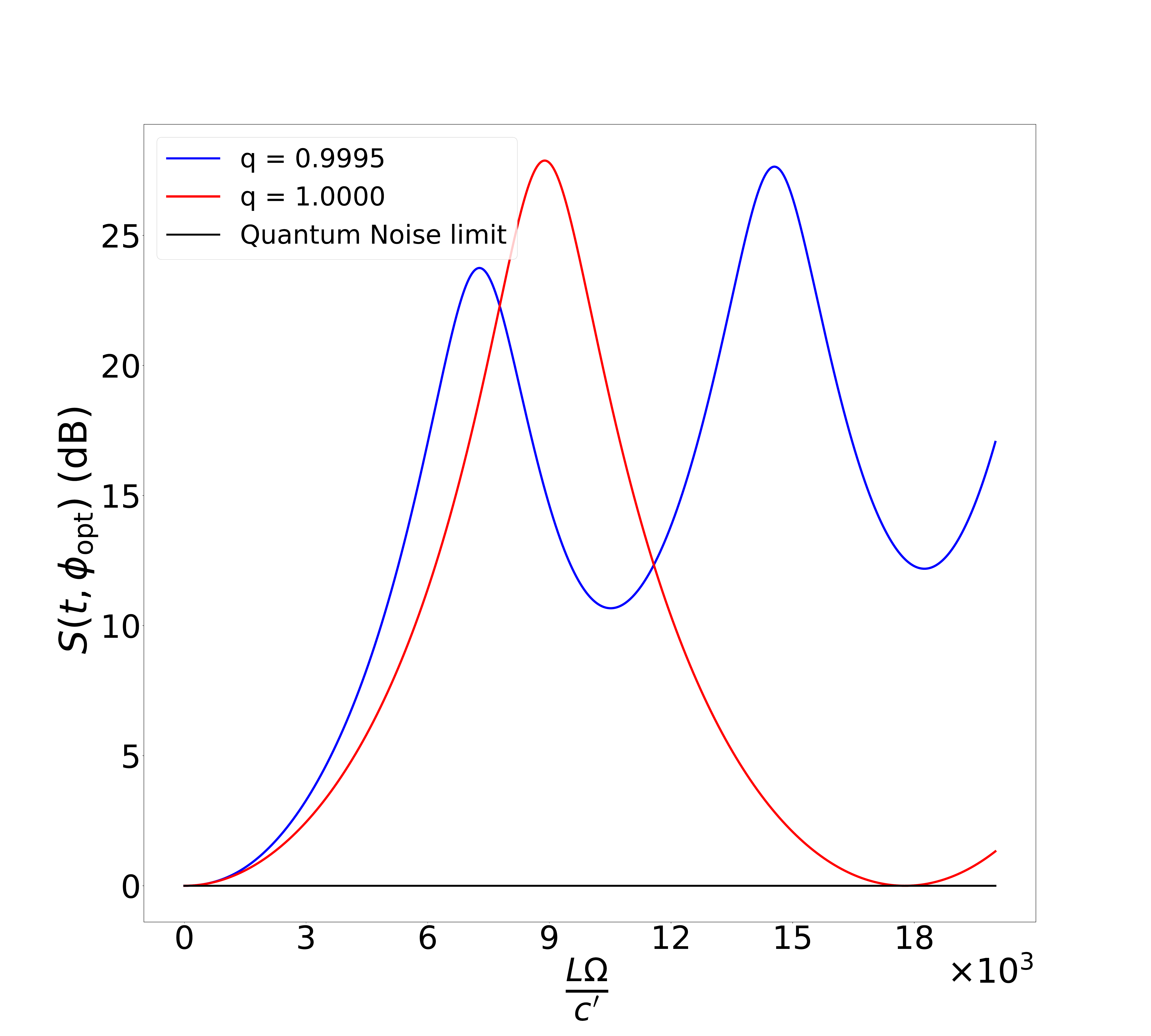}
\caption{Vacuum noise squeezing $S(t=L/c',\phi_{{\rm opt}})$ for optimum phase given by Eq.~(\ref{phiopt}) as a function of propagation length $L$ using parameters realistic for silicon on insulator waveguides. 
The parameters are $\omega_L/\Omega = 12.4$, $\eta = 10^{-3}$, and Raman shifts $q = 0.9995$ (blue) and $q = 1$ (red). The black line represents the quantum noise limit of 0 dB.}
\label{fig:squeezing_t}
\end{figure}

\section{Photon vacuum squeezing and correlation}


The Schrieffer-Wolff perturbative series predicts a two-mode squeezed state (\ref{psiSW}) away from the regime of phonon resonance ($q$ away from $1$), when no phonons are produced and the phonon occupation remains close to zero. However, it gives rise to a singularity as $q\rightarrow 1$. We now use our nonperturbative theory to predict the dynamics of the gas of photonic Cooper pairs in the regime of phonon resonance. This can be done by writing bare operators in terms of Ramaniton ones, and obtaining their Heisenberg representation from $\alpha_j(t)=e^{i{\cal H}_R t/\hbar}\alpha_j(0)e^{-i{\cal H}_R t/\hbar}=e^{-i\omega_j t}\alpha_j(0)$  \NM{(See Appendix~\ref{appendixNambu} for resulting expressions)}. 

One important constraint is that the Ramaniton model (\ref{RamanitonModel}) has the symmetry $[{\cal H}_R, N_S-N_{aS}-N_c]=0$, where $N_c=c^{\dag}c$ and $N_j=b^{\dag}_jb_j$ with $j=S,aS$ are bare phonon and photon number operators. As a result the quantity $\langle N_S\rangle-\langle N_{aS}\rangle-\langle N_c\rangle$ is a constant independent of time. 
If the initial state is the vacuum of the bare modes, $|\psi(t=0)\rangle= \ket{0}_B\equiv |0\rangle_S |0\rangle_{c} |0\rangle_{aS}$, their average occupation changes in time according to $\langle N_S(t)\rangle-\langle N_{aS}(t)\rangle=\langle N_c(t)\rangle$. Therefore, $\langle N_{S}(t)\rangle-\langle N_{aS}(t)\rangle$ oscillates as time evolves due to coherent population transfer between photons and phonons. This is shown in Fig.~\ref{fig:occupation}. 

When $\langle N_c(t)\rangle=0$ we have $\langle N_{S}(t)\rangle=\langle N_{aS}(t)\rangle$, a special situation where the photonic state is ``paired up''. As we now show this gives rise to maximum squeezing. \NM{For this reason, the condition $\langle N_c\rangle=0$ with $\langle N_S\rangle=\langle N_{aS}\rangle > 0$ will be denoted \emph{resonant squeezing}.}

Define the generalized quadrature \cite{Gerry2004}
\begin{equation}
\label{gen_quad}
X(t) = \frac{1}{2^{3/2}}\left\{e^ {-i\phi} \left[b_S(t) + b_{aS}(t)\right] + {\rm H.c.}\right\},
\end{equation}
where $b_j(t) = e^{i{\cal H}_R t/\hbar}b_j(0)e^{-i{\cal H}_R t/\hbar}$ are photon operators in the Heisenberg representation, and $\phi$ is a phase. The amount of vacuum noise squeezing as a function of time in Decibel units (dB) is given by
\begin{equation}
    S(t,\phi)=-10 \log_{10}\left\{
    \frac{\left\langle \left[\Delta X(t)\right]^{2}\right\rangle}{\left\langle \left[\Delta X(0)\right]^{2}\right\rangle}
    \right\},
    \label{Sdb}
\end{equation}
where $\Delta X(t)=X(t)-\langle X\rangle$. The averages are taken by assuming the initial state is the vacuum of the bare modes in the rotating frame,  $|\psi(t=0)\rangle= 
\ket{0}_B$.
The denominator in Eq.~(\ref{Sdb}) is the variance of $X$ in the vacuum, $\langle \left[\Delta X(0)\right]^{2}\rangle=1/4$. This reference corresponds to the quantum noise limit (QNL) of $0$~dB. 
The generalized quadrature is said to be squeezed (anti-squeezed) whenever the variance of $X$ is below (above) the QNL \cite{Gerry2004, Walls2008}.

\NM{Using Eqs.~(\ref{bSt}),~(\ref{baSt}) we obtain
\begin{eqnarray}
\frac{\left\langle \left[\Delta X(t)\right]^{2}\right\rangle}{\left\langle \left[\Delta X(0)\right]^{2}\right\rangle}
&=&\left|X_S(t)+e^{2i\phi}Y_{as}^{*}(t)\right|^{2}\nonumber\\
&=& 1+\left\langle N_S(t)\right\rangle+\left\langle N_{aS}(t)\right\rangle\nonumber\\
&&+\left[e^{-2i\phi}X_S(t)Y_{aS}(t)+{\rm c.c.}\right].
\end{eqnarray}
The noise ratio is minimized (maximum squeezing) when $e^{-2i\phi}X_S(t)Y_{aS}(t)$ is \emph{negative}, equal to $-\sqrt{1 + \langle N_S(t) \rangle}\sqrt{\langle N_{aS}(t)\rangle}$. This happens when the phase $\phi$ is equal to the optimal phase
\begin{equation}
    \phi_{\rm opt}(t)=\frac{1}{2}\left\{\pi +{\rm Arg}\left[X_S(t)Y_{aS}(t)\right]\right\}.
\label{phiopt}
\end{equation}
When $\phi=\phi_{{\rm opt}}$ we can write
\begin{equation}
    \frac{\left\langle \left[\Delta X(t)\right]^{2}\right\rangle}{\left\langle \left[\Delta X(0)\right]^{2}\right\rangle}\Biggr|_{\phi=\phi_{{\rm opt
    }}}\!\!\!\!\!\!\!\!\!\!\!\!
     = \left| \sqrt{1 + \langle N_S(t) \rangle} - \sqrt{\langle N_{S}(t) \rangle - \langle N_{c}(t) \rangle} \right|^2.
\end{equation}
This equation is minimized (maximum squeezing) when $\langle N_{c}(t)\rangle=0$ simultaneously with $\langle N_{S}(t)\rangle \rightarrow\infty$, i.e. when the brightness of the Stokes channel is maximized. 
The latter takes place at the phonon resonance, when $q=1$ and $\phi_{{\rm opt}}(t)=\pi/2$ is independent of time}.

\begin{figure}
\centering
\includegraphics[width = \linewidth]{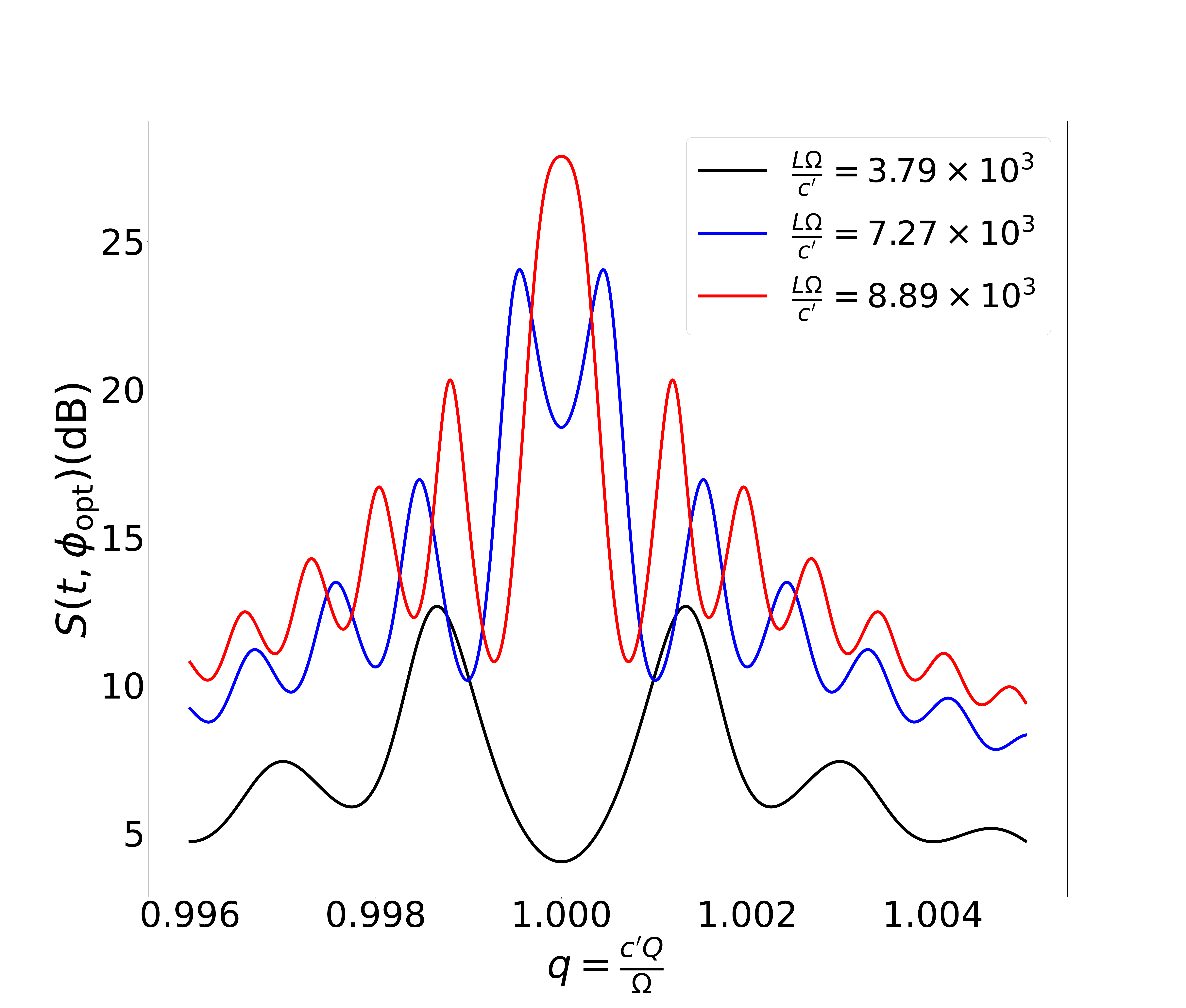}
\caption{Vacuum noise squeezing $S(t=L/c',\phi_{{\rm opt}})$ for realistic silicon waveguides as  a function of Raman shift $q$, for three different waveguide lengths $L$. Global maximum squeezing of 28~dB is achieved for $L=8.89\times 10^{3}c'/\Omega$ and $q=1$, the phonon resonance condition when a real phonon is emitted/absorbed. For other propagation lengths squeezing has local maxima away from $q=1$. The parameters used are $\omega_L/\Omega = 12.4$ and $\eta = 10^{-3}$. 
\label{dos}}
\end{figure}

Our calculations show that the amount of vacuum squeezing $S(t,\phi_{{\rm opt}})$ is oscillatory as a function of the propagation time $t$ or length $L=c't$ of a waveguide; this is shown in Fig.~{\ref{fig:squeezing_t}}, where $\Omega t$ is now replaced by $L/(c'/\Omega)$, with $L$ the length of a photonic waveguide. 
The quantum oscillations as a function of $L$ arise from the population exchange between bare phonon and photons in the time evolution of the Ramaniton. As a consequence, for each Raman shift $q$ there is a value of nonzero $t$ or $L$ that gives rise to $\langle N_c\rangle=0$ and maximum squeezing. 
Figure~\ref{dos} shows vacuum squeezing as a function of Raman shift $q$ for three different waveguide lengths $L$. The global maximum for squeezing is obtained for $L\Omega/c'=8.89\times 10^3$ and $q=1$, when both the condition for resonant squeezing ($\langle N_c\rangle =0$) and the condition for phonon resonance ($q=1$, emission/absorption of a real phonon) are \emph{simultaneously} satisfied. 
For other values of $L$, the condition for resonant squeezing $\langle N_c\rangle =0$ is satisfied away from $q=1$; as a result squeezing shows local maxima at these points.

\begin{figure}
\centering
\includegraphics[width = \linewidth]{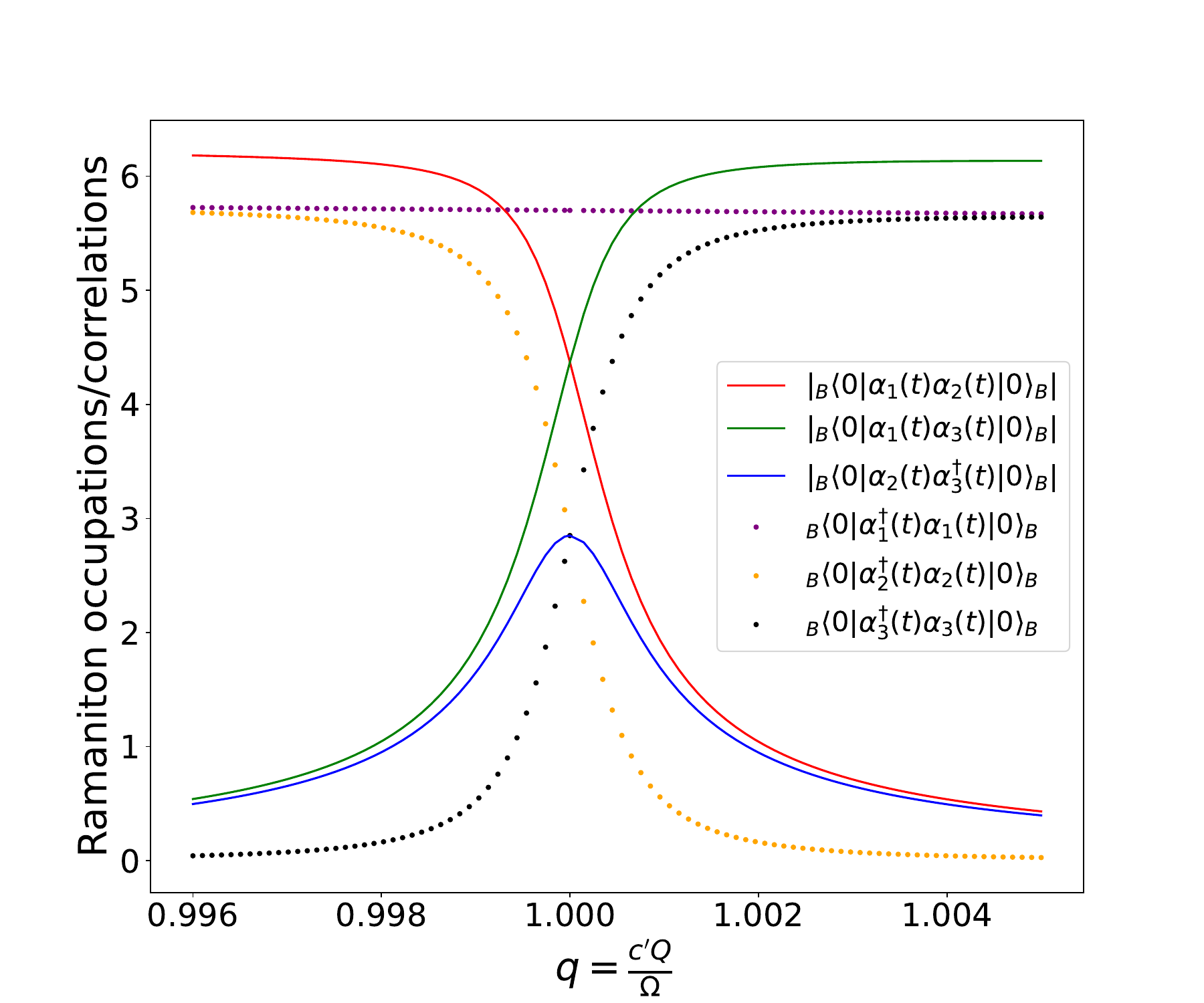}
\caption{\NM{Ramaniton occupations $\langle \alpha^{\dag}_{j}(t)\alpha_j(t)\rangle$ and modulus of correlations $|\langle\alpha_{j}(t)\alpha_k(t)\rangle|$ for the evolved bare vacuum as a function of $q$. Note that both are independent of time. At $q=1$ the sum of correlations is maximized, leading to maximum destructive interference at $t=t_{{\rm RS}}$ as showed in Eq.~(\ref{noiseq=1}).
\label{fig:occup_corr}}}
\end{figure}

\NM{The condition for resonant squeezing, $\langle N_c\rangle=0$ with $\langle N_S\rangle=\langle N_{aS}\rangle > 0$, requires the system to recur to a zero phonon state without recurring back to the full vacuum state $\ket{0}_B$. The reason why this can happen is related to the fact that the first Ramaniton is decoupled from the phonon, see Eq.~(\ref{alpha1bare}). As a result, the phonon  operator in the Heisenberg picture $c(t)$ is a linear combination of only two Ramanitons, $\alpha_{2}(t)$ and $\alpha_3(t)$ (See Eq.~(\ref{c})).  As these two Ramanitons evolve in time they will interfere constructively when their phase difference is the same, $e^{i(\omega_3-\omega_2) t}=1$. 
When this happens, $c(t)$ is the same as $c(0)$ apart from a global phase, and $\langle N_c(t)\rangle=0$. This shows that resonant squeezing occurs when $t=t_{{\rm RS}}$ for
\begin{equation}
    t_{{\rm RS}}=\frac{2\pi n_{q,\eta}}{\omega_3-\omega_2}=\frac{2\pi n_{q,\eta}}{\Omega\sqrt{\left(q-1\right)^{2}+\frac{\eta^2q}{2}}}, 
\end{equation}
with $n_{q,\eta}$ a positive integer that avoids full recurrence to $\langle N_S\rangle=\langle N_{aS}\rangle = 0$. Since the operators $b^{\dag}_S(t)$ and $ b_{aS}(t)$ both involve Ramaniton $\alpha_{1}^{\dag}(t)$ \emph{in addition} to $\alpha_{2}(t)$ and $\alpha_3(t)$, full recurrence requires $e^{i(\omega_3+\omega_1) t}=e^{i(\omega_3-\omega_2) t}=1$. But this can only happen if $(\omega_3+\omega_1)$ is commensurable with $(\omega_3-\omega_2)$
\begin{equation}
    n_{q,\eta}=\frac{\omega_3-\omega_2}{\omega_3+\omega_1}m,
\end{equation}
for a positive integer $m$. For $q=1$ we get $(\omega_3-\omega_2)/(\omega_3+\omega_1)=2$, showing that resonant squeezing occurs only for $n_{1,\eta}$=odd, i.e. $\Omega t_{{\rm RS}}=(2\sqrt{2}\pi/\eta) n_{1,\eta}$ with $n_{1,\eta}=1,3,5,\ldots$. At even multiples of $2\sqrt{2}\pi/\eta$ we get recurrence to $\ket{0}_B$. 
This behaviour is seen in Fig.~\ref{fig:occupation} for $\eta=10^{-3}$ and $2\sqrt{2}\pi/\eta=8.89 \times 10^{3}$.

The global maximum for squeezing can also be related to Ramaniton occupations and correlations. Figure~\ref{fig:occup_corr} shows the Ramaniton occupations $\langle \alpha^{\dag}_{j}(t)\alpha_j(t)\rangle$ and correlations $\langle\alpha_{j}(t)\alpha_k(t)\rangle$ for the evolved bare vacuum as a function of $q$. Note that the former is independent of time, and the latter oscillates as $e^{-i(\omega_j+\omega_k)t}$. At $q=1$ and $\phi=\phi_{{\rm opt}}=\pi/2$ the noise as a function of $t$ becomes
\begin{widetext}
\begin{eqnarray}
\label{Noise_t=tR_q =1}
     \frac{\left\langle \left[\Delta X(t)\right]^{2}\right\rangle}{\left\langle \left[\Delta X(0)\right]^{2}\right\rangle}\Biggr|_{\phi=\frac{\pi}{2}, q=1}\!\!\!\!\!\!\!\!\!\!\!\!
     &=& 
     \left[\left(\frac{\omega_L}{\Omega}\right)-\sqrt{\left(\frac{\omega_L}{\Omega}\right)^{2}-1}\right]
     \left\{1 + \langle \alpha_1^{\dag}(t)\alpha_1(t)\rangle+ \frac{1}{2}\left[\langle \alpha^{\dag}_{2}(t)\alpha_2(t)\rangle +\langle\alpha^{\dag}_{3}(t)\alpha_3(t)\rangle\right]\right.\nonumber\\
     &&\left.+ \left|\langle\alpha_{2}(t)\alpha^{\dag}_{3}(t)\rangle\right|\cos{\left[(\omega_3-\omega_2)t\right]}
     +\sqrt{2}\Big[\left|\langle \alpha_1(t)\alpha_2(t)\rangle\right|+\left|\langle \alpha_1(t)\alpha_3(t)\rangle\right|\Big]\cos{\left[(\omega_1+\omega_3)t\right]}
     \right\}.
     \label{noiseq=1}
\end{eqnarray}
\end{widetext}
%
When $t=t_{{\rm RS}}$, $\cos{\left[(\omega_3-\omega_2)t\right]}=1$, and $\cos{\left[(\omega_1+\omega_3)t\right]}=-1$, showing that the Ramaniton correlations $\langle\alpha_{j}(t)\alpha_k(t)\rangle$ interfere \emph{destructively} with the occupations. On top of this, Fig.~\ref{fig:occup_corr} shows that at $q=1$ the sum of correlations is maximized. The interference is maximally destructive, leading to the global minimum of the noise,
\begin{equation}
\frac{\left\langle \left[\Delta X(t_{{\rm RS}})\right]^{2}\right\rangle}{\left\langle \left[\Delta X(0)\right]^{2}\right\rangle}\Biggr|_{\phi=\frac{\pi}{2}, q=1}= \left\{\left(\frac{\omega_L}{\Omega}\right)-\sqrt{\left(\frac{\omega_L}{\Omega}\right)^{2}-1}\right\}^{2},
\label{opt_noise}
\end{equation}
which goes to zero when $\omega_L/\Omega\rightarrow \infty$}. 

\begin{figure}
    \centering
    \includegraphics[width = \linewidth]{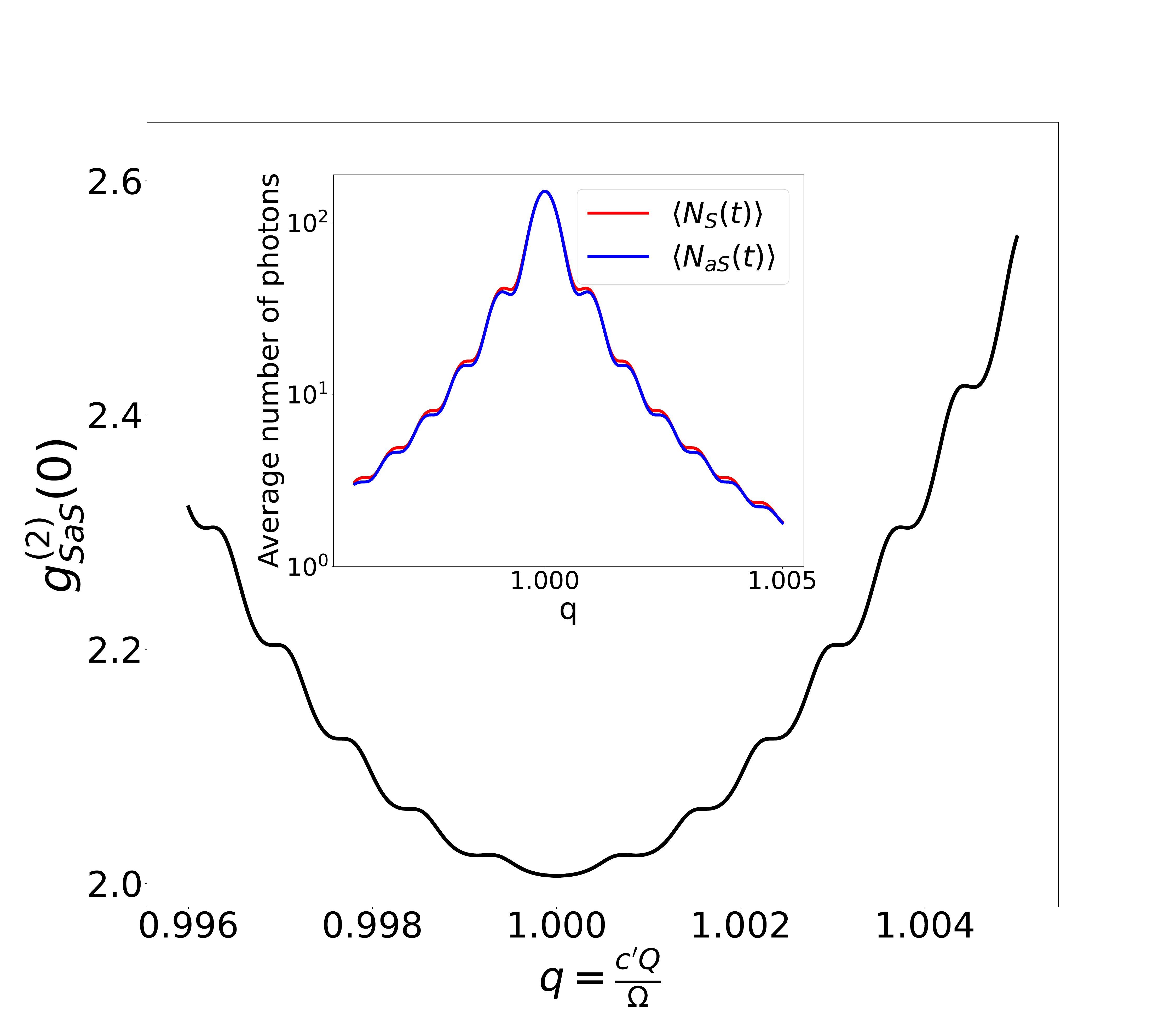}
    \caption{Two-mode intensity cross-correlation function $g^{(2)}_{SaS}(0)$ as a function of Raman shift $q$ for the parameters leading to a global maximum in squeezing.  This is waveguide length 
    $L\Omega/c'=8.89 \times 10^3$ and other parameters as in Fig.~\ref{dos}. The inset is the plot of average number of Stokes and anti-Stokes photons for the same set of parameters. 
    \label{g2_0}}
\end{figure}

For additional insight, we turn to calculations of the second-order cross-correlation between photons in S and aS modes. The zero-time delay two-mode intensity correlation function is defined as \cite{Walls2008}
\begin{equation}
    g^{(2)}_{SaS}(0) = \frac{\langle b_S^{\dagger}(t)b_{aS}^{\dagger}(t)b_{aS}(t) b_{S}(t)\rangle}{\langle b_S^{\dagger}(t)b_S(t)\rangle \langle b_{aS}^{\dagger}(t)b_{aS}(t)\rangle}. 
    \label{g2}
\end{equation}
When the initial state is the vacuum, we can use the Nambu method to obtain the exact result 
\begin{equation}
g^{(2)}_{SaS}(0) = 2 +\frac{1}{\langle N_S(t) \rangle}.
\label{g2exact}
\end{equation}
This shows that in the Ramaniton model $g^{(2)}_{SaS}(0)$ is a proxy for the number of photons in the S mode. Figure~\ref{g2_0} shows $g^{(2)}_{SaS}(0)$ as a function of $q$ for $L=8.89\times 10^{3} c'/\Omega$ and other parameters the same as in Fig.~\ref{dos}. We see that $g^{(2)}_{SaS}(0)$ is minimum at the global maximum squeezing point at $q=1$. For this case $g^{(2)}_{SaS}(0)$ \emph{anticorrelates} with the amount of squeezing.

The inset of Fig.~\ref{g2_0} shows the number of S and aS photons as a function of $q$ for the same parameters. Away from $q=1$, $\langle N_S(t)\rangle\approx \langle N_{aS}(t)\rangle$ decreases and as a result $g^{(2)}_{SaS}(0)$ increases, see Eq.~(\ref{g2exact}). 
When $q\ll 1$ or $q\gg 1$, both $\langle N_S(t)\rangle$ and $\langle N_{aS}(t)\rangle$ are quite small, leading to $g^{(2)}_{SaS}(0)\gg 2$ (not shown). 

\section{Conclusions}

We presented a nonperturbative theory for phonon-mediated photon-photon interaction based on the concept of the Ramaniton, an excitation that is qualitatively different than the
polariton in that it mixes particle and antiparticle (hole-like) phonon/photon modes. This distinctive feature arises from Cooper-pair-like correlations that are \emph{universal} to all Raman scattering processes. 
The Ramaniton removes the distinction between real and virtual phonon processes in correlated Raman scattering, regularizing the singularity that appears in the perturbative treatment for this problem \cite{Saraiva2017}. 

\NM{While exciton-polaritons are also known to mix photon excitations with anti-exciton ones, the amount of mixing is antiresonant, in that it can be safely neglected in the rotating wave approximation \cite{Hopfield1958, Ciuti2005}. Thus in most circumstances exciton-polaritons nearly conserve the number of excitations and do not show anomalous quantum effects. A notable exception occurs in Fabry-Perot cavities with multiple quantum wells, where ultra-strong light-exciton coupling amplifies the quantum nature of the exciton-polariton. In this regime, interesting quantum effects related to squeezing and photon pair generation are reported to occur \cite{Ciuti2005}. 

We remark that the Ramaniton is qualitatively different from exciton-polaritons, in that its admixture between photons and antiphotons  remains strong after taking the rotating wave approximation (\ref{RamanitonModel}) and even become ``resonant'' at $q=1$. From Eqs.~(\ref{alpha1bare}),~(\ref{alpha2}),~and~(\ref{alpha3}) the amount of photon/antiphoton admixture is $\eta_{-}/\eta_{+}\sim 1$ for all $q$. Thus, our quantum theory of Raman scattering reveals opportunities for the demonstration of quantum electrodynamical effects with conventional materials and devices}.

The usefulness of the Ramaniton is evident from Fig.~\ref{fig:occupation}. As time evolves, the number of photons/phonons goes from $0$ to $\sim 10^{2}$; capturing this range of modes with exact numerical diagonalization is challenging. The Ramaniton solves this problem and enables analytical determination of the regimes where optimal squeezing takes place. 

We show that the amount of two-mode vacuum noise squeezing, a property that is well-known to arise from entanglement \cite{Duan2000} shows local maxima (resonances) as a function of photon time of flight $t$ or waveguide propagation length $L=c't$. ``Resonant squeezing'' occurs when the average number of excited phonons $\langle N_c\rangle$ becomes equal to zero \NM{at the same time that the number of Stokes and antiStokes photons is nonzero}. When this happens the number of Stokes and antiStokes photons become equal, leading to ``photonic pairing'', with squeezing measured in dB increasing logarithmically with the number of photons. 

The global maximum for squeezing happens when both the condition for resonant squeezing $\langle N_c\rangle =0$ and the condition for phonon resonance $q=1$ are simultaneously satisfied. In this case the ``paired state'' has maximum brightness \NM{and the amount of squeezing increases logarithmically with the laser frequency as shown in Eq.~(\ref{opt_noise})}.
As seen in the $q=1$ curve of Fig.~\ref{fig:squeezing_t}, the global maximum peak is relatively broad as a function of $L$, so maximum squeezing can be achieved without the need of high precision on the value of $L$. This leads to our prediction that up to 28~dB of squeezing may be reached in silicon on insulator waveguides, provided that light can be extracted near the optimal waveguide length $L=8.89\times 10^{3}c'/\Omega = 7.95$~mm. Achieving this would be a significant improvement on the 15~dB of squeezing measured in an optical table \cite{Vahlbruch2016} and also on the 11~dB of squeezing inferred before extraction on optical fibers \cite{Dong2008} and photonic chips \cite{Nehra2022}. 

We expect similar resonant squeezing phenomena to occur in resonators that trap photons for extended periods of time. The photon trapping time $\tau$ will play a similar role as the propagation time $t=L/c'$ in waveguides, leading to optimal squeezing  for certain values of $\tau$ that depend not only on resonator length but also on resonator quality factor $Q$.

The conventional mechanism for generating squeezed states in waveguides is to use photon evolution under the Kerr effect in order to convert a coherent state into a one-mode squeezed state \cite{Andersen2016}. This process is optimized in optical fibers measuring several meters long, and it was pointed out that Raman scattering out of the squeezed photon mode provides the limiting factor for maximum squeezing \cite{Dong2008}. Our work presents a qualitatively different mechanism, where Raman scattering acts instead as the generator of two-mode squeezed states, leading to maximum squeezing in waveguides with much shorter length. 

In conclusion, we described a theory for phonon-mediated generation of squeezed light. The theory opens up avenues for material and device 
optimization in integrated quantum photonics for quantum information and sensing applications. 

\section*{Acknowledgements}
We thank S. Barzanjeh, J. Fabian, B. King, and A. MacRae for encouragement and insightful discussions in this topic. This work was supported by the Natural Sciences and Engineering Research Council of Canada (NSERC) through its Discovery (RGPIN-2020-04328), CREATE (543245-2020), and USRA programs. We acknowledge the support of the Government of Canada’s New Frontiers in Research Fund (NFRF), [NFRFE-2019-01104].

\appendix

\NM{\section{Exact diagonalization of the Ramaniton Hamiltonian with the Nambu method \label{appendixNambu}}

The Nambu method provides a sistematic way to carry out the Bogoliubov transformation necessary to put the Ramaniton Hamiltonian in diagonal form (\ref{diagonal-ham}). In the Nambu representation the Hamiltonian~(\ref{RamanitonModel}) is written as
$\mathcal{H}_R = \frac{1}{2} \bm{v}^{\dagger} \cdot \mathbb{L} \cdot \bm{v}$,
where $\bm{v} = \left(b_S, c, b_{aS}, b_S^{\dagger}, c^{\dagger}, b_{aS}^{\dagger}\right)^T$ is a particle-antiparticle column vector and $\mathbb{L}$ is a $6 \times 6$ Hermitian matrix. Our goal is to find the $6\times 6$ nonunitary transformation $\mathbb{U}$ that connects $\bm{v}$ to the Ramaniton operators $\bm{\alpha}$, as in $\bm{v}=\mathbb{U}\cdot\bm{\alpha}$.

As explained in section~\ref{sec:nonpert} of the paper, the columns of $\mathbb{U}$ are formed by the eigenvectors of $\mathbb{Z \cdot L}$, where $\mathbb{Z}={\rm diag}\{1,1,1,-1,-1,-1 \}$ is the plus (minus) identity in the particle (antiparticle) subspace. Inspection of Eq.~(\ref{RamanitonModel}) leads to
\begin{equation}
    \mathbb{Z}\cdot \mathbb{L} = \hbar\Omega\begin{pmatrix}
        -q & 0 & 0 & 0 & i\eta_- & 0 \\
         0 & 1 & -i \eta_+ & i\eta_- & 0 & 0 \\
         0 & i \eta_+ & q & 0 & 0 & 0 \\
         0 & i \eta_- & 0 & q & 0 & 0 \\
         i\eta_- & 0 & 0 & 0 & -1 & -i\eta_+ \\
         0 & 0 & 0 & 0 & i\eta_+ & -q \\
         
    \end{pmatrix}.
\end{equation}
We observe that 
\begin{equation}
    \mathbb{Z}\cdot \mathbb{L} \begin{pmatrix}
        \xi_1 \\
        0\\
        0\\
        0\\
        \xi_2\\
        \xi_3
    \end{pmatrix} = \hbar\Omega\begin{pmatrix}
        -q \xi_1 + i \eta_- \xi_2 \\
        0\\
        0\\
        0\\
        i\eta_- \xi_1 - \xi_2 - i\eta_+ \xi_3\\
        i\eta_+ \xi_2 - q \xi_3
    \end{pmatrix},
\end{equation}
showing that $\mathbb{Z}\cdot \mathbb{L}$ is block diagonal. The first block is given by
\begin{equation}
    T = \Omega\begin{pmatrix}
        -q & i \eta_- & 0 \\
        i\eta_- &-1 &-i\eta_+ \\
        0 & i\eta_+ & -q
    \end{pmatrix},
\end{equation}
with eigenvalues
\begin{subequations}
\begin{align}
        \lambda_1 & = -q\Omega, \\
        \lambda_2 &= \left(- \frac{q+1}{2} + \frac{1}{2}\sqrt{(q - 1)^2 + \frac{\eta^2q}{2}}\right)\Omega, \\
        \lambda_3 &=  \left(- \frac{q+1}{2} - \frac{1}{2}\sqrt{(q - 1)^2 + \frac{\eta^2q}{2}}\right)\Omega.
\end{align}
\end{subequations}
The eigenvector of matrix $T$ corresponding to the eigenvalue $\lambda_1$ is
\begin{subequations}
\label{eigvec_w1}
\begin{align}
    \xi_1^{(1)} & =  \frac{\eta_+}{\sqrt{\eta_+^2 - \eta_-^2}}, \\
    \xi_2^{(1)} & =  0 ,\\
    \xi_3^{(1)} & = \frac{\eta_-}{\sqrt{\eta_+^2 - \eta_-^2}}.
\end{align}    
\end{subequations}
Note that this eigenvector can be chosen to satisfy the positive normalization rule $\left|\xi^{(1)}_{1}\right|^{2}-\left|\xi^{(1)}_{2}\right|^{2}-\left|\xi^{(1)}_{3}\right|^{2}=+1$,  so it can be chosen as the first column of $\mathbb{U}$. Thus the first Ramaniton has frequency $\omega_1=\lambda_1$. Recall that the normalization condition for the columns of $\mathbb{U}$ are set by the requirement that the transformation is canonical, $\left(\mathbb{Z}\cdot \mathbb{U}^{\dag}\cdot \mathbb{Z}\right)\cdot \mathbb{U}=\mathbb{I}$. Hence the first three columns of $\mathbb{U}$ must satisfy the positive normalization rule, with the other three satisfying the negative rule described below. 

The remaining eigenvectors of matrix $T$ corresponding to eigenvalues $\lambda_j$ $(j = 2, 3)$ are 
\begin{subequations}
\label{eigvec_w23}
    \begin{align}
        \xi_1^{(j)}  = &  \frac{i\eta_-}{\sqrt{(q + \frac{\lambda_j}{\Omega})^2 + \eta_+^2 - \eta_-^2}}, \\
        \xi_2^{(j)}  = & \frac{\left(q + \frac{\lambda_j}{\Omega}\right)}{\sqrt{(q + \frac{\lambda_j}{\Omega})^2 + \eta_+^2 - \eta_-^2}}, \\
        \xi_3^{(j)}  = &  \frac{i\eta_+}{\sqrt{(q + \frac{\lambda_j}{\Omega})^2 + \eta_+^2 - \eta_-^2}}.
    \end{align}
\end{subequations}
Because $\eta_{+}>\eta_{-}$, these \emph{can not} satisfy the positive normalization rule; instead they have to be normalized by the negative rule, $\left|\xi^{(j)}_{1}\right|^{2}-\left|\xi^{(j)}_{2}\right|^{2}-\left|\xi^{(j)}_{3}\right|^{2}=-1$. So we have to choose these two remaining eigenvectors to be the fifth and sixth columns of $\mathbb{U}$, respectively. Their associated eigenvalues correspond to negative Ramaniton frequencies, so we must have $\omega_2=-\lambda_2$ and $\omega_3=-\lambda_3$ for the remaining two Ramanitons. These Ramaniton eigenfrequencies are the ones displayed in Eqs.~(\ref{omega1})-(\ref{omega3}).

The other $3\times 3$ block of $\mathbb{Z}\cdot\mathbb{L}$ is
\begin{equation}
    \widetilde{T} = \Omega
    \begin{pmatrix}
        1 & -i\eta_+ & i \eta_- \\
        i\eta_+ & q &0 \\
        i \eta_- & 0 & q
    \end{pmatrix},
\end{equation}
with eigenvalues $\tilde{\lambda}_j=-\lambda_j$ and eigenvectors $(\xi_{2}^{(j)*},\xi_{3}^{(j)*},\xi_{1}^{(j)*})^{T}$.
The $j=2,3$ eigenvectors satisfy the positive normalization rule, so they belong to the second and third columns of $\mathbb{U}$, respectively. The $j=1$ eigenvector satisfies negative normalization so it forms the fourth column of $\mathbb{U}$. 

Putting all eigenvectors together we get
\begin{equation}
    \mathbb{U} =
    \begin{pmatrix}
        \xi_{1}^{(1)} & 0 & 0 & 0 & \xi_{1}^{(2)} & \xi_{1}^{(3)} \\
        0 & \xi_{2}^{(2)*} & \xi_{2}^{(3)*} & \xi_{2}^{(1)*} & 0 & 0 \\
        0 & \xi_{3}^{(2)*} & \xi_{3}^{(3)*} & \xi_{3}^{(1)*} & 0 & 0 \\
        0 & \xi_{1}^{(2)*} & \xi_{1}^{(3)*} & \xi_{1}^{(1)*} & 0 & 0 \\
        \xi_{2}^{(1)} & 0 & 0 & 0 & \xi_{2}^{(2)} & \xi_{2}^{(3)} \\
        \xi_{3}^{(1)} & 0 & 0 & 0 & \xi_{3}^{(2)} & \xi_{3}^{(3)} \\        
    \end{pmatrix}.
\end{equation}
This implies
\begin{eqnarray}
    b_S &=& \xi_{1}^{(1)}\alpha_1 +\xi_{1}^{(2)}\alpha_{2}^{\dag} +\xi_{1}^{(3)}\alpha_{3}^{\dag},\label{bS}\\
    c &=& \xi_{2}^{(2)*}\alpha_2 +\xi_{2}^{(3)*}\alpha_3,\label{c}\\
    b_{aS} &=& \xi_{3}^{(2)*}\alpha_2 +\xi_{3}^{(3)*}\alpha_3, +\xi_{3}^{(1)*}\alpha_{1}^{\dag},\label{baS}    
\end{eqnarray}
and the inverse relations
\begin{eqnarray}
    \alpha_1 &=& \xi_{1}^{(1)*}b_S -\xi_{3}^{(1)}b_{aS}^{\dag},\label{alpha1}\\
    \alpha_2 &=& -\xi_{1}^{(2)}b_{S}^{\dag} +\xi_{2}^{(2)}c+\xi_{3}^{(2)}b_{aS},\label{alpha2}\\
    \alpha_3 &=& -\xi_{1}^{(3)}b_{S}^{\dag} +\xi_{2}^{(3)}c+\xi_{3}^{(3)}b_{aS}.\label{alpha3}
\end{eqnarray}
These are used to find the time evolution of photon operators in the Heisenberg representation. Plug $\alpha_j(t)=e^{-i\omega_jt}\alpha_j(0)$ into Eqs.~(\ref{bS})-(\ref{baS}) and use Eqs.~(\ref{alpha1})-(\ref{alpha3}) to express $\alpha_j(0)$ in terms of bare photons at $t=0$. The result is
\begin{eqnarray}
    b_S(t)&=&X_S(t)b_S(0)+Y_S(t)b_{aS}^{\dag}(0)\nonumber\\
    &&+Z_S(t)c^{\dag}(0),\label{bSt}\\
    b_{aS}(t)&=&X_{aS}(t)b_{aS}(0)+Y_{aS}(t)b_{S}^{\dag}(0)\nonumber\\
    &&+Z_{aS}(t)c(0),\label{baSt}
\end{eqnarray}
with for example 
\begin{eqnarray}
    X_S(t)&=&\left|\xi_{1}^{(1)}\right|^{2}e^{-i\omega_1t}-\left|\xi_{1}^{(2)}\right|^{2}e^{i\omega_2t}\nonumber\\
    &&-\left|\xi_{1}^{(3)}\right|^{2}e^{i\omega_3t},\label{X_St}\\
    Y_{aS}(t)&=&-\xi_{3}^{(2)*}\xi_{1}^{(2)}e^{-i\omega_2t}-\xi_{3}^{(3)*}\xi_{1}^{(3)}e^{-i\omega_3t}\nonumber\\
    &&+\xi_{3}^{(1)*}\xi_{1}^{(1)}e^{i\omega_1t}.\label{Y_aSt}
\end{eqnarray}
Using Eqs.~(\ref{bSt})~and~(\ref{baSt}) we obtain $_B\langle 0|N_{S}(t)|0\rangle_B=\left|Y_{S}(t)\right|^{2}+\left|Z_{S}(t)\right|^{2}=\left|X_{S}(t)\right|^{2}-1$ (the last identity comes from $[b_S(t),b^{\dag}_{S}(t)]=1$), and $_B\langle 0|N_{aS}(t)|0\rangle_B=\left|Y_{aS}(t)\right|^{2}$. 

}

\bibliography{SaS}

\begin{thebibliography}{28}%
\makeatletter
\providecommand \@ifxundefined [1]{%
 \@ifx{#1\undefined}
}%
\providecommand \@ifnum [1]{%
 \ifnum #1\expandafter \@firstoftwo
 \else \expandafter \@secondoftwo
 \fi
}%
\providecommand \@ifx [1]{%
 \ifx #1\expandafter \@firstoftwo
 \else \expandafter \@secondoftwo
 \fi
}%
\providecommand \natexlab [1]{#1}%
\providecommand \enquote  [1]{``#1''}%
\providecommand \bibnamefont  [1]{#1}%
\providecommand \bibfnamefont [1]{#1}%
\providecommand \citenamefont [1]{#1}%
\providecommand \href@noop [0]{\@secondoftwo}%
\providecommand \href [0]{\begingroup \@sanitize@url \@href}%
\providecommand \@href[1]{\@@startlink{#1}\@@href}%
\providecommand \@@href[1]{\endgroup#1\@@endlink}%
\providecommand \@sanitize@url [0]{\catcode `\\12\catcode `\$12\catcode
  `\&12\catcode `\#12\catcode `\^12\catcode `\_12\catcode `\%12\relax}%
\providecommand \@@startlink[1]{}%
\providecommand \@@endlink[0]{}%
\providecommand \url  [0]{\begingroup\@sanitize@url \@url }%
\providecommand \@url [1]{\endgroup\@href {#1}{\urlprefix }}%
\providecommand \urlprefix  [0]{URL }%
\providecommand \Eprint [0]{\href }%
\providecommand \doibase [0]{https://doi.org/}%
\providecommand \selectlanguage [0]{\@gobble}%
\providecommand \bibinfo  [0]{\@secondoftwo}%
\providecommand \bibfield  [0]{\@secondoftwo}%
\providecommand \translation [1]{[#1]}%
\providecommand \BibitemOpen [0]{}%
\providecommand \bibitemStop [0]{}%
\providecommand \bibitemNoStop [0]{.\EOS\space}%
\providecommand \EOS [0]{\spacefactor3000\relax}%
\providecommand \BibitemShut  [1]{\csname bibitem#1\endcsname}%
\let\auto@bib@innerbib\@empty
\bibitem [{\citenamefont {Wang}\ \emph {et~al.}(2020)\citenamefont {Wang},
  \citenamefont {Sciarrino}, \citenamefont {Laing},\ and\ \citenamefont
  {Thompson}}]{Wang2020}%
  \BibitemOpen
  \bibfield  {author} {\bibinfo {author} {\bibfnamefont {J.}~\bibnamefont
  {Wang}}, \bibinfo {author} {\bibfnamefont {F.}~\bibnamefont {Sciarrino}},
  \bibinfo {author} {\bibfnamefont {A.}~\bibnamefont {Laing}},\ and\ \bibinfo
  {author} {\bibfnamefont {M.~G.}\ \bibnamefont {Thompson}},\ }\bibfield
  {title} {\bibinfo {title} {Integrated photonic quantum technologies},\ }\href
  {https://doi.org/10.1038/s41566-019-0532-1} {\bibfield  {journal} {\bibinfo
  {journal} {Nat. Photon.}\ }\textbf {\bibinfo {volume} {14}},\ \bibinfo
  {pages} {273} (\bibinfo {year} {2020})}\BibitemShut {NoStop}%
\bibitem [{\citenamefont {Schnabel}(2017)}]{Schnabel2017}%
  \BibitemOpen
  \bibfield  {author} {\bibinfo {author} {\bibfnamefont {R.}~\bibnamefont
  {Schnabel}},\ }\bibfield  {title} {\bibinfo {title} {Squeezed states of light
  and their applications in laser interferometers},\ }\href
  {https://doi.org/10.1016/j.physrep.2017.04.001} {\bibfield  {journal}
  {\bibinfo  {journal} {Phys. Rep.}\ }\textbf {\bibinfo {volume} {684}},\
  \bibinfo {pages} {1} (\bibinfo {year} {2017})}\BibitemShut {NoStop}%
\bibitem [{\citenamefont {Fukui}\ \emph {et~al.}(2018)\citenamefont {Fukui},
  \citenamefont {Tomita}, \citenamefont {Okamoto},\ and\ \citenamefont
  {Fujii}}]{Fukui2018}%
  \BibitemOpen
  \bibfield  {author} {\bibinfo {author} {\bibfnamefont {K.}~\bibnamefont
  {Fukui}}, \bibinfo {author} {\bibfnamefont {A.}~\bibnamefont {Tomita}},
  \bibinfo {author} {\bibfnamefont {A.}~\bibnamefont {Okamoto}},\ and\ \bibinfo
  {author} {\bibfnamefont {K.}~\bibnamefont {Fujii}},\ }\bibfield  {title}
  {\bibinfo {title} {High-threshold fault-tolerant quantum computation with
  analog quantum error correction},\ }\href
  {https://doi.org/10.1103/PhysRevX.8.021054} {\bibfield  {journal} {\bibinfo
  {journal} {Phys. Rev. X}\ }\textbf {\bibinfo {volume} {8}},\ \bibinfo {pages}
  {021054} (\bibinfo {year} {2018})}\BibitemShut {NoStop}%
\bibitem [{\citenamefont {Wehner}\ \emph {et~al.}(2018)\citenamefont {Wehner},
  \citenamefont {Elkouss},\ and\ \citenamefont {Hanson}}]{Wehner2018}%
  \BibitemOpen
  \bibfield  {author} {\bibinfo {author} {\bibfnamefont {S.}~\bibnamefont
  {Wehner}}, \bibinfo {author} {\bibfnamefont {D.}~\bibnamefont {Elkouss}},\
  and\ \bibinfo {author} {\bibfnamefont {R.}~\bibnamefont {Hanson}},\
  }\bibfield  {title} {\bibinfo {title} {Quantum internet: A vision for the
  road ahead},\ }\href {https://doi.org/10.1126/science.aam9288} {\bibfield
  {journal} {\bibinfo  {journal} {Science}\ }\textbf {\bibinfo {volume}
  {362}},\ \bibinfo {pages} {eaam9288} (\bibinfo {year} {2018})}\BibitemShut
  {NoStop}%
\bibitem [{\citenamefont {Duan}\ \emph {et~al.}(2000)\citenamefont {Duan},
  \citenamefont {Giedke}, \citenamefont {Cirac},\ and\ \citenamefont
  {Zoller}}]{Duan2000}%
  \BibitemOpen
  \bibfield  {author} {\bibinfo {author} {\bibfnamefont {L.-M.}\ \bibnamefont
  {Duan}}, \bibinfo {author} {\bibfnamefont {G.}~\bibnamefont {Giedke}},
  \bibinfo {author} {\bibfnamefont {J.~I.}\ \bibnamefont {Cirac}},\ and\
  \bibinfo {author} {\bibfnamefont {P.}~\bibnamefont {Zoller}},\ }\bibfield
  {title} {\bibinfo {title} {Inseparability criterion for continuous variable
  systems},\ }\href {https://doi.org/10.1103/PhysRevLett.84.2722} {\bibfield
  {journal} {\bibinfo  {journal} {Phys. Rev. Lett.}\ }\textbf {\bibinfo
  {volume} {84}},\ \bibinfo {pages} {2722} (\bibinfo {year}
  {2000})}\BibitemShut {NoStop}%
\bibitem [{\citenamefont {Kasperczyk}\ \emph {et~al.}(2016)\citenamefont
  {Kasperczyk}, \citenamefont {de~Aguiar~J\'unior}, \citenamefont {Rabelo},
  \citenamefont {Saraiva}, \citenamefont {Santos}, \citenamefont {Novotny},\
  and\ \citenamefont {Jorio}}]{Kasperczyk2016}%
  \BibitemOpen
  \bibfield  {author} {\bibinfo {author} {\bibfnamefont {M.}~\bibnamefont
  {Kasperczyk}}, \bibinfo {author} {\bibfnamefont {F.~S.}\ \bibnamefont
  {de~Aguiar~J\'unior}}, \bibinfo {author} {\bibfnamefont {C.}~\bibnamefont
  {Rabelo}}, \bibinfo {author} {\bibfnamefont {A.}~\bibnamefont {Saraiva}},
  \bibinfo {author} {\bibfnamefont {M.~F.}\ \bibnamefont {Santos}}, \bibinfo
  {author} {\bibfnamefont {L.}~\bibnamefont {Novotny}},\ and\ \bibinfo {author}
  {\bibfnamefont {A.}~\bibnamefont {Jorio}},\ }\bibfield  {title} {\bibinfo
  {title} {Temporal quantum correlations in inelastic light scattering from
  water},\ }\href {https://doi.org/10.1103/PhysRevLett.117.243603} {\bibfield
  {journal} {\bibinfo  {journal} {Phys. Rev. Lett.}\ }\textbf {\bibinfo
  {volume} {117}},\ \bibinfo {pages} {243603} (\bibinfo {year}
  {2016})}\BibitemShut {NoStop}%
\bibitem [{\citenamefont {Velez}\ \emph {et~al.}(2020)\citenamefont {Velez},
  \citenamefont {Sudhir}, \citenamefont {Sangouard},\ and\ \citenamefont
  {Galland}}]{Velez2020}%
  \BibitemOpen
  \bibfield  {author} {\bibinfo {author} {\bibfnamefont {S.~T.}\ \bibnamefont
  {Velez}}, \bibinfo {author} {\bibfnamefont {V.}~\bibnamefont {Sudhir}},
  \bibinfo {author} {\bibfnamefont {N.}~\bibnamefont {Sangouard}},\ and\
  \bibinfo {author} {\bibfnamefont {C.}~\bibnamefont {Galland}},\ }\bibfield
  {title} {\bibinfo {title} {Bell correlations between light and vibration at
  ambient conditions},\ }\href {https://doi.org/10.1126/sciadv.abb0260}
  {\bibfield  {journal} {\bibinfo  {journal} {Sci. Adv.}\ }\textbf {\bibinfo
  {volume} {6}},\ \bibinfo {pages} {eabb0260} (\bibinfo {year}
  {2020})}\BibitemShut {NoStop}%
\bibitem [{\citenamefont {Saraiva}\ \emph {et~al.}(2017)\citenamefont
  {Saraiva}, \citenamefont {de~Aguiar~J\'unior}, \citenamefont {de~Melo~e
  Souza}, \citenamefont {Pena}, \citenamefont {Monken}, \citenamefont {Santos},
  \citenamefont {Koiller},\ and\ \citenamefont {Jorio}}]{Saraiva2017}%
  \BibitemOpen
  \bibfield  {author} {\bibinfo {author} {\bibfnamefont {A.}~\bibnamefont
  {Saraiva}}, \bibinfo {author} {\bibfnamefont {F.~S.}\ \bibnamefont
  {de~Aguiar~J\'unior}}, \bibinfo {author} {\bibfnamefont {R.}~\bibnamefont
  {de~Melo~e Souza}}, \bibinfo {author} {\bibfnamefont {A.~P.}\ \bibnamefont
  {Pena}}, \bibinfo {author} {\bibfnamefont {C.~H.}\ \bibnamefont {Monken}},
  \bibinfo {author} {\bibfnamefont {M.~F.}\ \bibnamefont {Santos}}, \bibinfo
  {author} {\bibfnamefont {B.}~\bibnamefont {Koiller}},\ and\ \bibinfo {author}
  {\bibfnamefont {A.}~\bibnamefont {Jorio}},\ }\bibfield  {title} {\bibinfo
  {title} {{Photonic Counterparts of Cooper Pairs}},\ }\href
  {https://doi.org/10.1103/PhysRevLett.119.193603} {\bibfield  {journal}
  {\bibinfo  {journal} {Phys. Rev. Lett.}\ }\textbf {\bibinfo {volume} {119}},\
  \bibinfo {pages} {193603} (\bibinfo {year} {2017})}\BibitemShut {NoStop}%
\bibitem [{\citenamefont {Guimar\~aes}\ \emph {et~al.}(2020)\citenamefont
  {Guimar\~aes}, \citenamefont {Santos}, \citenamefont {Jorio},\ and\
  \citenamefont {Monken}}]{Guimaraes2020}%
  \BibitemOpen
  \bibfield  {author} {\bibinfo {author} {\bibfnamefont {A.~V.~A.}\
  \bibnamefont {Guimar\~aes}}, \bibinfo {author} {\bibfnamefont {M.~F.}\
  \bibnamefont {Santos}}, \bibinfo {author} {\bibfnamefont {A.}~\bibnamefont
  {Jorio}},\ and\ \bibinfo {author} {\bibfnamefont {C.~H.}\ \bibnamefont
  {Monken}},\ }\bibfield  {title} {\bibinfo {title} {Stokes--anti-stokes
  light-scattering process: A photon-wave-function approach},\ }\href
  {https://doi.org/10.1103/PhysRevA.102.033719} {\bibfield  {journal} {\bibinfo
   {journal} {Phys. Rev. A}\ }\textbf {\bibinfo {volume} {102}},\ \bibinfo
  {pages} {033719} (\bibinfo {year} {2020})}\BibitemShut {NoStop}%
\bibitem [{\citenamefont {Boyd}(2020)}]{Boyd2020}%
  \BibitemOpen
  \bibfield  {author} {\bibinfo {author} {\bibfnamefont {R.~W.}\ \bibnamefont
  {Boyd}},\ }\href@noop {} {\emph {\bibinfo {title} {Nonlinear Optics}}}\
  (\bibinfo  {publisher} {Academic Press},\ \bibinfo {address} {London, U.K.},\
  \bibinfo {year} {2020})\ Chap.~\bibinfo {chapter} {5}\BibitemShut {NoStop}%
\bibitem [{\citenamefont {Banic}\ \emph {et~al.}(2022)\citenamefont {Banic},
  \citenamefont {Zatti}, \citenamefont {Liscidini},\ and\ \citenamefont
  {Sipe}}]{Banic2022}%
  \BibitemOpen
  \bibfield  {author} {\bibinfo {author} {\bibfnamefont {M.}~\bibnamefont
  {Banic}}, \bibinfo {author} {\bibfnamefont {L.}~\bibnamefont {Zatti}},
  \bibinfo {author} {\bibfnamefont {M.}~\bibnamefont {Liscidini}},\ and\
  \bibinfo {author} {\bibfnamefont {J.~E.}\ \bibnamefont {Sipe}},\ }\bibfield
  {title} {\bibinfo {title} {Two strategies for modeling nonlinear optics in
  lossy integrated photonic structures},\ }\href
  {https://doi.org/10.1103/PhysRevA.106.043707} {\bibfield  {journal} {\bibinfo
   {journal} {Phys. Rev. A}\ }\textbf {\bibinfo {volume} {106}},\ \bibinfo
  {pages} {043707} (\bibinfo {year} {2022})}\BibitemShut {NoStop}%
\bibitem [{\citenamefont {Kittel}(1987)}]{Kittel1987}%
  \BibitemOpen
  \bibfield  {author} {\bibinfo {author} {\bibfnamefont {C.}~\bibnamefont
  {Kittel}},\ }\href@noop {} {\emph {\bibinfo {title} {Quantum Theory of
  Solids}}},\ \bibinfo {edition} {2nd}\ ed.\ (\bibinfo  {publisher} {Wiley},\
  \bibinfo {address} {New York},\ \bibinfo {year} {1987})\ Chap.~\bibinfo
  {chapter} {3}\BibitemShut {NoStop}%
\bibitem [{\citenamefont {Diniz}\ and\ \citenamefont
  {de~Sousa}(2020)}]{Diniz2020}%
  \BibitemOpen
  \bibfield  {author} {\bibinfo {author} {\bibfnamefont {I.}~\bibnamefont
  {Diniz}}\ and\ \bibinfo {author} {\bibfnamefont {R.}~\bibnamefont
  {de~Sousa}},\ }\bibfield  {title} {\bibinfo {title} {{Intrinsic Photon Loss
  at the Interface of Superconducting Devices}},\ }\href
  {https://doi.org/10.1103/PhysRevLett.125.147702} {\bibfield  {journal}
  {\bibinfo  {journal} {Phys. Rev. Lett.}\ }\textbf {\bibinfo {volume} {125}},\
  \bibinfo {pages} {147702} (\bibinfo {year} {2020})}\BibitemShut {NoStop}%
\bibitem [{\citenamefont {Jackson}(1998)}]{Jackson1998}%
  \BibitemOpen
  \bibfield  {author} {\bibinfo {author} {\bibfnamefont {J.~D.}\ \bibnamefont
  {Jackson}},\ }\href@noop {} {\emph {\bibinfo {title} {Classical
  Electrodynamics}}},\ \bibinfo {edition} {3rd}\ ed.\ (\bibinfo  {publisher}
  {Wiley},\ \bibinfo {address} {New York},\ \bibinfo {year} {1998})\
  Chap.~\bibinfo {chapter} {12}\BibitemShut {NoStop}%
\bibitem [{\citenamefont {Gerry}\ and\ \citenamefont
  {Knight}(2004)}]{Gerry2004}%
  \BibitemOpen
  \bibfield  {author} {\bibinfo {author} {\bibfnamefont {C.}~\bibnamefont
  {Gerry}}\ and\ \bibinfo {author} {\bibfnamefont {P.}~\bibnamefont {Knight}},\
  }\href {https://doi.org/10.1017/CBO9780511791239} {\emph {\bibinfo {title}
  {Introductory Quantum Optics}}}\ (\bibinfo  {publisher} {Cambridge University
  Press},\ \bibinfo {address} {Cambridge, U.K.},\ \bibinfo {year} {2004})\
  Chap.~\bibinfo {chapter} {7}\BibitemShut {NoStop}%
\bibitem [{\citenamefont {Walls}\ and\ \citenamefont
  {Milburn}(2008)}]{Walls2008}%
  \BibitemOpen
  \bibfield  {author} {\bibinfo {author} {\bibfnamefont {D.~F.}\ \bibnamefont
  {Walls}}\ and\ \bibinfo {author} {\bibfnamefont {G.~J.}\ \bibnamefont
  {Milburn}},\ }\href@noop {} {\emph {\bibinfo {title} {Quantum Optics}}},\
  \bibinfo {edition} {2nd}\ ed.\ (\bibinfo  {publisher} {Springer-Verlag},\
  \bibinfo {address} {Berlin, Germany},\ \bibinfo {year} {2008})\
  Chap.~\bibinfo {chapter} {5}\BibitemShut {NoStop}%
\bibitem [{\citenamefont {Nambu}(1960)}]{Nambu1960}%
  \BibitemOpen
  \bibfield  {author} {\bibinfo {author} {\bibfnamefont {Y.}~\bibnamefont
  {Nambu}},\ }\bibfield  {title} {\bibinfo {title} {Quasi-particles and gauge
  invariance in the theory of superconductivity},\ }\href
  {https://doi.org/10.1103/PhysRev.117.648} {\bibfield  {journal} {\bibinfo
  {journal} {Phys. Rev.}\ }\textbf {\bibinfo {volume} {117}},\ \bibinfo {pages}
  {648} (\bibinfo {year} {1960})}\BibitemShut {NoStop}%
\bibitem [{\citenamefont {Beairsto}\ \emph {et~al.}(2021)\citenamefont
  {Beairsto}, \citenamefont {Cazayous}, \citenamefont {Fishman},\ and\
  \citenamefont {de~Sousa}}]{Beairsto2021}%
  \BibitemOpen
  \bibfield  {author} {\bibinfo {author} {\bibfnamefont {S.}~\bibnamefont
  {Beairsto}}, \bibinfo {author} {\bibfnamefont {M.}~\bibnamefont {Cazayous}},
  \bibinfo {author} {\bibfnamefont {R.~S.}\ \bibnamefont {Fishman}},\ and\
  \bibinfo {author} {\bibfnamefont {R.}~\bibnamefont {de~Sousa}},\ }\bibfield
  {title} {\bibinfo {title} {Confined magnons},\ }\href
  {https://doi.org/10.1103/PhysRevB.104.134415} {\bibfield  {journal} {\bibinfo
   {journal} {Phys. Rev. B}\ }\textbf {\bibinfo {volume} {104}},\ \bibinfo
  {pages} {134415} (\bibinfo {year} {2021})}\BibitemShut {NoStop}%
\bibitem [{\citenamefont {Hart}\ \emph {et~al.}(1970)\citenamefont {Hart},
  \citenamefont {Aggarwal},\ and\ \citenamefont {Lax}}]{Hart1970}%
  \BibitemOpen
  \bibfield  {author} {\bibinfo {author} {\bibfnamefont {T.~R.}\ \bibnamefont
  {Hart}}, \bibinfo {author} {\bibfnamefont {R.~L.}\ \bibnamefont {Aggarwal}},\
  and\ \bibinfo {author} {\bibfnamefont {B.}~\bibnamefont {Lax}},\ }\bibfield
  {title} {\bibinfo {title} {Temperature dependence of raman scattering in
  silicon},\ }\href {https://doi.org/10.1103/PhysRevB.1.638} {\bibfield
  {journal} {\bibinfo  {journal} {Phys. Rev. B}\ }\textbf {\bibinfo {volume}
  {1}},\ \bibinfo {pages} {638} (\bibinfo {year} {1970})}\BibitemShut {NoStop}%
\bibitem [{\citenamefont {Schinke}\ \emph {et~al.}(2015)\citenamefont
  {Schinke}, \citenamefont {Christian~Peest}, \citenamefont {Schmidt},
  \citenamefont {Brendel}, \citenamefont {Bothe}, \citenamefont {Vogt},
  \citenamefont {Kröger}, \citenamefont {Winter}, \citenamefont {Schirmacher},
  \citenamefont {Lim}, \citenamefont {Nguyen},\ and\ \citenamefont
  {MacDonald}}]{Schinke2015}%
  \BibitemOpen
  \bibfield  {author} {\bibinfo {author} {\bibfnamefont {C.}~\bibnamefont
  {Schinke}}, \bibinfo {author} {\bibfnamefont {P.}~\bibnamefont
  {Christian~Peest}}, \bibinfo {author} {\bibfnamefont {J.}~\bibnamefont
  {Schmidt}}, \bibinfo {author} {\bibfnamefont {R.}~\bibnamefont {Brendel}},
  \bibinfo {author} {\bibfnamefont {K.}~\bibnamefont {Bothe}}, \bibinfo
  {author} {\bibfnamefont {M.~R.}\ \bibnamefont {Vogt}}, \bibinfo {author}
  {\bibfnamefont {I.}~\bibnamefont {Kröger}}, \bibinfo {author} {\bibfnamefont
  {S.}~\bibnamefont {Winter}}, \bibinfo {author} {\bibfnamefont
  {A.}~\bibnamefont {Schirmacher}}, \bibinfo {author} {\bibfnamefont
  {S.}~\bibnamefont {Lim}}, \bibinfo {author} {\bibfnamefont {H.~T.}\
  \bibnamefont {Nguyen}},\ and\ \bibinfo {author} {\bibfnamefont
  {D.}~\bibnamefont {MacDonald}},\ }\bibfield  {title} {\bibinfo {title}
  {{Uncertainty analysis for the coefficient of band-to-band absorption of
  crystalline silicon}},\ }\href {https://doi.org/10.1063/1.4923379} {\bibfield
   {journal} {\bibinfo  {journal} {AIP Advances}\ }\textbf {\bibinfo {volume}
  {5}},\ \bibinfo {pages} {067168} (\bibinfo {year} {2015})}\BibitemShut
  {NoStop}%
\bibitem [{\citenamefont {Leuthold}\ \emph {et~al.}(2010)\citenamefont
  {Leuthold}, \citenamefont {Koos},\ and\ \citenamefont
  {Freude}}]{Leuthold2010}%
  \BibitemOpen
  \bibfield  {author} {\bibinfo {author} {\bibfnamefont {J.}~\bibnamefont
  {Leuthold}}, \bibinfo {author} {\bibfnamefont {C.}~\bibnamefont {Koos}},\
  and\ \bibinfo {author} {\bibfnamefont {W.}~\bibnamefont {Freude}},\
  }\bibfield  {title} {\bibinfo {title} {Nonlinear silicon photonics},\ }\href
  {https://doi.org/10.1038/nphoton.2010.185} {\bibfield  {journal} {\bibinfo
  {journal} {Nat. Photon.}\ }\textbf {\bibinfo {volume} {4}},\ \bibinfo {pages}
  {535} (\bibinfo {year} {2010})}\BibitemShut {NoStop}%
\bibitem [{\citenamefont {Vaidya}\ \emph {et~al.}(2020)\citenamefont {Vaidya},
  \citenamefont {Morrison}, \citenamefont {Helt}, \citenamefont {Shahrokshahi},
  \citenamefont {Mahler}, \citenamefont {Collins}, \citenamefont {Tan},
  \citenamefont {Lavoie}, \citenamefont {Repingon}, \citenamefont {Menotti},
  \citenamefont {Quesada}, \citenamefont {Pooser}, \citenamefont {Lita},
  \citenamefont {Gerrits}, \citenamefont {Nam},\ and\ \citenamefont
  {Vernon}}]{Vaidya2020}%
  \BibitemOpen
  \bibfield  {author} {\bibinfo {author} {\bibfnamefont {V.}~\bibnamefont
  {Vaidya}}, \bibinfo {author} {\bibfnamefont {B.}~\bibnamefont {Morrison}},
  \bibinfo {author} {\bibfnamefont {L.}~\bibnamefont {Helt}}, \bibinfo {author}
  {\bibfnamefont {R.}~\bibnamefont {Shahrokshahi}}, \bibinfo {author}
  {\bibfnamefont {D.}~\bibnamefont {Mahler}}, \bibinfo {author} {\bibfnamefont
  {M.}~\bibnamefont {Collins}}, \bibinfo {author} {\bibfnamefont
  {K.}~\bibnamefont {Tan}}, \bibinfo {author} {\bibfnamefont {J.}~\bibnamefont
  {Lavoie}}, \bibinfo {author} {\bibfnamefont {A.}~\bibnamefont {Repingon}},
  \bibinfo {author} {\bibfnamefont {M.}~\bibnamefont {Menotti}}, \bibinfo
  {author} {\bibfnamefont {N.}~\bibnamefont {Quesada}}, \bibinfo {author}
  {\bibfnamefont {R.}~\bibnamefont {Pooser}}, \bibinfo {author} {\bibfnamefont
  {A.}~\bibnamefont {Lita}}, \bibinfo {author} {\bibfnamefont {T.}~\bibnamefont
  {Gerrits}}, \bibinfo {author} {\bibfnamefont {S.~W.}\ \bibnamefont {Nam}},\
  and\ \bibinfo {author} {\bibfnamefont {Z.}~\bibnamefont {Vernon}},\
  }\bibfield  {title} {\bibinfo {title} {Broadband quadrature-squeezed vacuum
  and nonclassical photon number correlations from a nanophotonic device},\
  }\href {https://doi.org/10.1126/sciadv.aba9186} {\bibfield  {journal}
  {\bibinfo  {journal} {Sci. Adv.}\ }\textbf {\bibinfo {volume} {6}},\ \bibinfo
  {pages} {eaba9186} (\bibinfo {year} {2020})}\BibitemShut {NoStop}%
\bibitem [{\citenamefont {Hopfield}(1958)}]{Hopfield1958}%
  \BibitemOpen
  \bibfield  {author} {\bibinfo {author} {\bibfnamefont {J.~J.}\ \bibnamefont
  {Hopfield}},\ }\bibfield  {title} {\bibinfo {title} {Theory of the
  contribution of excitons to the complex dielectric constant of crystals},\
  }\href {https://doi.org/10.1103/PhysRev.112.1555} {\bibfield  {journal}
  {\bibinfo  {journal} {Phys. Rev.}\ }\textbf {\bibinfo {volume} {112}},\
  \bibinfo {pages} {1555} (\bibinfo {year} {1958})}\BibitemShut {NoStop}%
\bibitem [{\citenamefont {Ciuti}\ \emph {et~al.}(2005)\citenamefont {Ciuti},
  \citenamefont {Bastard},\ and\ \citenamefont {Carusotto}}]{Ciuti2005}%
  \BibitemOpen
  \bibfield  {author} {\bibinfo {author} {\bibfnamefont {C.}~\bibnamefont
  {Ciuti}}, \bibinfo {author} {\bibfnamefont {G.}~\bibnamefont {Bastard}},\
  and\ \bibinfo {author} {\bibfnamefont {I.}~\bibnamefont {Carusotto}},\
  }\bibfield  {title} {\bibinfo {title} {Quantum vacuum properties of the
  intersubband cavity polariton field},\ }\href
  {https://doi.org/10.1103/PhysRevB.72.115303} {\bibfield  {journal} {\bibinfo
  {journal} {Phys. Rev. B}\ }\textbf {\bibinfo {volume} {72}},\ \bibinfo
  {pages} {115303} (\bibinfo {year} {2005})}\BibitemShut {NoStop}%
\bibitem [{\citenamefont {Vahlbruch}\ \emph {et~al.}(2016)\citenamefont
  {Vahlbruch}, \citenamefont {Mehmet}, \citenamefont {Danzmann},\ and\
  \citenamefont {Schnabel}}]{Vahlbruch2016}%
  \BibitemOpen
  \bibfield  {author} {\bibinfo {author} {\bibfnamefont {H.}~\bibnamefont
  {Vahlbruch}}, \bibinfo {author} {\bibfnamefont {M.}~\bibnamefont {Mehmet}},
  \bibinfo {author} {\bibfnamefont {K.}~\bibnamefont {Danzmann}},\ and\
  \bibinfo {author} {\bibfnamefont {R.}~\bibnamefont {Schnabel}},\ }\bibfield
  {title} {\bibinfo {title} {Detection of 15 db squeezed states of light and
  their application for the absolute calibration of photoelectric quantum
  efficiency},\ }\href {https://doi.org/10.1103/PhysRevLett.117.110801}
  {\bibfield  {journal} {\bibinfo  {journal} {Phys. Rev. Lett.}\ }\textbf
  {\bibinfo {volume} {117}},\ \bibinfo {pages} {110801} (\bibinfo {year}
  {2016})}\BibitemShut {NoStop}%
\bibitem [{\citenamefont {Dong}\ \emph {et~al.}(2008)\citenamefont {Dong},
  \citenamefont {Heersink}, \citenamefont {Corney}, \citenamefont {Drummond},
  \citenamefont {Andersen},\ and\ \citenamefont {Leuchs}}]{Dong2008}%
  \BibitemOpen
  \bibfield  {author} {\bibinfo {author} {\bibfnamefont {R.}~\bibnamefont
  {Dong}}, \bibinfo {author} {\bibfnamefont {J.}~\bibnamefont {Heersink}},
  \bibinfo {author} {\bibfnamefont {J.~f.}\ \bibnamefont {Corney}}, \bibinfo
  {author} {\bibfnamefont {P.~D.}\ \bibnamefont {Drummond}}, \bibinfo {author}
  {\bibfnamefont {U.~L.}\ \bibnamefont {Andersen}},\ and\ \bibinfo {author}
  {\bibfnamefont {G.}~\bibnamefont {Leuchs}},\ }\bibfield  {title} {\bibinfo
  {title} {{Experimental evidence for Raman-induced limits to efficient
  squeezing in optical fibers}},\ }\href {https://doi.org/10.1364/OL.33.000116}
  {\bibfield  {journal} {\bibinfo  {journal} {Optics letters}\ }\textbf
  {\bibinfo {volume} {33}},\ \bibinfo {pages} {116} (\bibinfo {year}
  {2008})}\BibitemShut {NoStop}%
\bibitem [{\citenamefont {Nehra}\ \emph {et~al.}(2022)\citenamefont {Nehra},
  \citenamefont {Sekine}, \citenamefont {Ledezma}, \citenamefont {Guo},
  \citenamefont {Gray}, \citenamefont {Roy},\ and\ \citenamefont
  {Marandi}}]{Nehra2022}%
  \BibitemOpen
  \bibfield  {author} {\bibinfo {author} {\bibfnamefont {R.}~\bibnamefont
  {Nehra}}, \bibinfo {author} {\bibfnamefont {R.}~\bibnamefont {Sekine}},
  \bibinfo {author} {\bibfnamefont {L.}~\bibnamefont {Ledezma}}, \bibinfo
  {author} {\bibfnamefont {Q.}~\bibnamefont {Guo}}, \bibinfo {author}
  {\bibfnamefont {R.~M.}\ \bibnamefont {Gray}}, \bibinfo {author}
  {\bibfnamefont {A.}~\bibnamefont {Roy}},\ and\ \bibinfo {author}
  {\bibfnamefont {A.}~\bibnamefont {Marandi}},\ }\bibfield  {title} {\bibinfo
  {title} {few-cycle vacuum squeezing in nanophotonics},\ }\href
  {https://doi.org/10.1126/science.abo6213} {\bibfield  {journal} {\bibinfo
  {journal} {Science}\ }\textbf {\bibinfo {volume} {377}},\ \bibinfo {pages}
  {1333} (\bibinfo {year} {2022})}\BibitemShut {NoStop}%
\bibitem [{\citenamefont {Andersen}\ \emph {et~al.}(2016)\citenamefont
  {Andersen}, \citenamefont {Gehring}, \citenamefont {Marquardt},\ and\
  \citenamefont {Leuchs}}]{Andersen2016}%
  \BibitemOpen
  \bibfield  {author} {\bibinfo {author} {\bibfnamefont {U.~L.}\ \bibnamefont
  {Andersen}}, \bibinfo {author} {\bibfnamefont {T.}~\bibnamefont {Gehring}},
  \bibinfo {author} {\bibfnamefont {C.}~\bibnamefont {Marquardt}},\ and\
  \bibinfo {author} {\bibfnamefont {G.}~\bibnamefont {Leuchs}},\ }\bibfield
  {title} {\bibinfo {title} {30 years of squeezed light generation},\ }\href
  {https://doi.org/10.1088/0031-8949/91/5/053001} {\bibfield  {journal}
  {\bibinfo  {journal} {Physica Scripta}\ }\textbf {\bibinfo {volume} {91}},\
  \bibinfo {pages} {053001} (\bibinfo {year} {2016})}\BibitemShut {NoStop}%
\end{thebibliography}%

\end{document}